\documentclass[aps,pra,twocolumn,a4paper,bibnotes]{revtex4-1}

\usepackage{tabularx}
\usepackage{array}
\usepackage{graphicx}
\usepackage{mathtools}
\usepackage{enumitem}

\usepackage{amsmath}
\usepackage{amssymb}
\usepackage{amsfonts}

\usepackage{phfqit}

\usepackage{mathptmx}

\usepackage{xcolor}
\definecolor{docnotelinkcolor}{rgb}{0,0,0.4}
\usepackage[bookmarks=true,backref=false]{hyperref}%
\hypersetup{unicode=true, %
  bookmarksnumbered=false,bookmarksopen=false,bookmarksopenlevel=1,%
  breaklinks=true,pdfborder={0 0 0},colorlinks=true}%
\hypersetup{%
  anchorcolor=docnotelinkcolor,citecolor=docnotelinkcolor, %
  filecolor=docnotelinkcolor,linkcolor=docnotelinkcolor, %
  menucolor=docnotelinkcolor,runcolor=docnotelinkcolor, %
  urlcolor=docnotelinkcolor}%

\newcommand{\heading}[1]{\par\vspace{1ex}\emph{#1.---}\ignorespaces}

\begin{document}

\title{Practical, Reliable Error Bars in Quantum Tomography}
\author{Philippe Faist}
\email{pfaist@phys.ethz.ch}
\affiliation{Institute for Theoretical Physics, ETH Zurich, 8093 Switzerland}
\author{Renato Renner}
\affiliation{Institute for Theoretical Physics, ETH Zurich, 8093 Switzerland}
\date{\today}

\begin{abstract}
  Precise characterization of quantum devices is usually achieved with quantum tomography.
  However, most methods which are currently widely used in experiments, such as maximum
  likelihood estimation, lack a well-justified error analysis. Promising recent methods
  based on confidence regions are difficult to apply in practice or yield error bars which
  are unnecessarily large.  Here, we propose a practical yet robust method for obtaining
  error bars.  We do so by introducing a novel representation of the output of the
  tomography procedure, the \emph{quantum error bars}.  This representation is (i)
  concise, being given in terms of few parameters, (ii) intuitive, providing a fair idea
  of the ``spread'' of the error, and (iii) useful, containing the necessary information
  for constructing confidence regions.  The statements resulting from our method are
  formulated in terms of a figure of merit, such as the fidelity to a reference state.  We
  present an algorithm for computing this representation and provide ready-to-use software.
  Our procedure is applied to actual experimental data obtained from two superconducting
  qubits in an entangled state, demonstrating the applicability of our method.
\end{abstract}

\maketitle

\heading{Introduction}
Recent experimental developments have demonstrated increasingly precise manipulation and
control of quantum systems, paving the way towards the hopeful implementation of a quantum
computer~\cite{Bennett2000_qitcomputation,Home2013AAMOP_ions,Devoret2013Sci_superconducting,Northup2014NatPhot_transfer,Riedel2010Nat_AtomChip,Monz2011PRL_14qubit,Maurer2012Sci_second,Usmani2012NPhot,Fedorov2012Nat,Steffen2013Nat_deterministic,Bussieres2014NPhot_teleportation,Harty2014PRL_highfidelity}.
The successful outcome of an experiment is usually certified using quantum tomography.
This is the task of inferring the quantum state of a device from statistics of
measurements on many copies of the
system~\cite{Helstrom1969_detection,Vogel1989PRA_tomo,DAriano2000PLA_universal,Cassinelli2000_group,Thew2002_PhysRevA66,DAriano2003AIEP_qt,BookParisRehacek2004_QuantumStateEstimation}.
Several methods perform this task and are widely used, such as maximum likelihood
estimation~\cite{Hradil1997PRA_MLE,Rehacek2001PRA}.

In the realistic regime where finite data are collected, the error bars provided by most
methods which are  widely applied in current
experiments~\cite{BookEfron1994Bootstrap,James2001PRA_KwiatQubitTomo,BookParisRehacek2004_QuantumStateEstimation,Home2009Sci_iontraps}
are typically ill justified and may lead to deceiving
conclusions~\cite{BlumeKohout2010_optimal,Jungnitsch2010PRL_increasing,BlumeKohout2012arXiv_Tomo}.
To remedy this problem, Blume-Kohout~\cite{BlumeKohout2012arXiv_Tomo} and Christandl and
Renner~\cite{Christandl2012_Tomo} resort to \emph{confidence regions}.  These are regions
in state space of all density matrices in which the state lies with high probability.  In
contrast to Bayesian methods~\cite{BlumeKohout2010_optimal}, the reliability statements do
not depend on any prior distribution.  However, confidence regions are \emph{a priori}
difficult to construct explicitly~\cite{Arrazola2013_reliable}.  Furthermore, they are
designed for worst-case scenarios and are often not representative of the intuitive extent
of the error.

Our main result is a novel representation of the output of the tomography procedure---a
summary of what the tomographic data tells us about the state of the system---which we
call \emph{quantum error bars}.  This description is (i)~concise, being given in terms of a
few parameters only, (ii)~intuitive, providing a fair idea of the ``spread'' of the error,
and (iii)~useful for precise statements, containing all necessary information for constructing
confidence regions.  Our method, in particular, inherits the mathematical robustness of the
confidence region approach.

The quantum error bars are designed to mimic the role of classical error bars.
Classically, an error bar typically represents the standard deviation of the distribution
of a physical quantity caused by noise or statistical errors; this distribution is usually
assumed to be Gaussian.  Observe that, precisely, classical error bars (i)~are a concise
description of the error, (ii)~provide a fair, intuitive idea of the ``spread'' of the
quantity of interest, and (iii)~allow us to calculate precise statements such as the
required error interval to consider (e.g., 5 standard deviations) for a specific
requested certainty level (e.g., one in a million).

Our statements are formulated in terms of a figure of merit which can be chosen freely.
Our method works best when the figure of merit is the fidelity to a pure target state, the
expectation value of an observable, or the trace distance to any reference state.  This
encompasses most tomography settings.

The quantum error bars are constructed as follows.  The input is the experimental data
from a general quantum tomography experiment.  Then we construct a particular
distribution $\mu(f)$ of the chosen figure of merit $f$, which has the property of
containing the necessary information to construct confidence regions at any confidence
level using the method of Ref.~\cite{Christandl2012_Tomo}.  We show that in a wide range
of situations and for a class of figures of merit, the distribution $\mu(f)$ can be
approximated by a simple analytical expression with three parameters.  The quantum
error bars are then straightforwardly deduced from these parameters.

We provide a simple numerical algorithm to obtain the quantum error bars from the
measurement data.  By fitting a numerical approximation to $\mu(f)$ with our approximate analytical model, we
obtain the values of the parameters of the model which directly translate to the quantum error bars.
The practicality of our method is demonstrated by applying it to experimental data from
two superconducting qubits.

Our work complements a vast literature which has provided error analyses for
experiments~\cite{Kiesel2008PRA_experimental,Rehacek2008NJP_diagnostics,Kiesel2010PRA_nonclassicality,BlumeKohout2010_verification,BlumeKohout2010PRL_HedgedMLE,Sugiyama2011PRA_Errors,Ferrie2012AIPConfProc,Rosset2012_implications,Shang2013_optimal,Walter2014IEEETIT_tomo_bounds,Sugiyama2014PRA_precision,Schwemmer2015PRL_systematic,Ball2016PRA_randomized}
as well as explicit
schemes~\cite{Lvovsky2004JOptB,BlumeKohout2006arXiv_honest,Schmied2011NJP_wigner,Dobek2011PRL_extraction,Smolin2012PRL_EfficientMLE,Anis2012NJP_coherent,Gill2013_asymptotic,Kech2015JPA_topology,Haack2015arXiv_MPS,Carpentier2015arXiv_uncertainty},
by introducing the novel concept of quantum error bars.  The complexity of such schemes
have also been investigated~\cite{Haah2015arXiv_sampleoptimal,ODonnell2015arXiv_efficient}
and numerical techniques put
forward~\cite{Rehacek2007PRA,BlumeKohout2010_optimal,Ferrie2015NJP_wrong,Granade2016NJP_bayesian}.
Furthermore, a number of contributions propose measurement schemes for fidelity
estimation~\cite{Silva2011PRL_practical,Flammia2011PRL_direct}, tomography of matrix product
states~\cite{Cramer2010_efficient}, estimation of low-rank
states~\cite{Gross2010PRL,Flammia2012NJP_compressed}, and permutationally invariant
tomography~\cite{Toth2010PRL_permutationally,Moroder2012NJP_permutationally_reconstr,Schwemmer2014PRL_comparison}.
An experiment following such schemes would achieve target benchmarks more efficiently, and it
could still be analyzed using our procedure, the latter being applicable to any
measurements.

The rest of this Letter is structured as follows. First, we briefly explain our quantum
tomography setup and the concept of confidence regions.  We then derive our main technical
results, namely, the definition of $\mu(f)$, its approximate theoretical model, and the
algorithm to estimate $\mu(f)$ numerically.  Finally, we demonstrate the applicability of our
method on experimental data.

\begin{figure}
  \centering%
  \includegraphics[width=86mm]{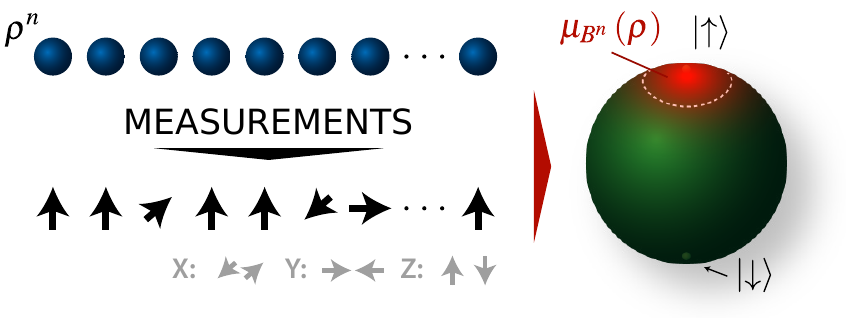}%
  \vspace*{-2mm}%
  \caption{Setup of quantum tomography. Measurements are taken on $n$ copies of a quantum
    system.  The outcomes allow us to infer what the state of the quantum system is. In this
    example a qubit is measured using Pauli operators. Here, the experimental data are
    most consistent with the state being $\ket\uparrow$, located at the top of the Bloch
    sphere (in green). However, because only finite data are collected, there is an
    uncertainty associated with this statement. In the method of
    Ref.~\cite{Christandl2012_Tomo}, a distribution
    $\mu_{B^n}\mathopen{}\left(\rho\right)$ (the red gradient) is determined from the data,
    from which confidence regions can be constructed (delimited by the dotted line). These are
    regions in state space in which the state lies with high probability.}
  \label{fig:TomoSetup}
\end{figure}

\heading{Quantum Tomography Setup}
A large number $n$ of copies of a quantum system are measured using independent, possibly
different, measurement settings (\autoref{fig:TomoSetup})~\footnote{In the general case considered 
in Ref.~\cite{Christandl2012_Tomo}, the measurements
  need not be independent, and the underlying quantum state of the $n$ systems only
  needs to be permutation-invariant. We focus on the  case of independent measurements for
  clarity of presentation and since this situation is the most widespread, although are
  results are expected to hold in the general case as well.}.
We list all the of the distinct positive operator-valued measure (POVM) effects in one set
$\{ E_k \}$, and denote by $n_k$ the
number of times the POVM effect $E_k$ was observed.  We then construct the likelihood
function, which will be needed in our analysis.  It is defined as the probability with
which the observed data would occur if the true state were $n$ copies of $\rho$,
\begin{align}
  \label{eq:Lambda}
  \Lambda(\rho) = \Pr\bigl[\mathrm{observed~data} \mid \rho\bigr]
  = \prod_k \bigl(\tr[E_k\rho]\bigr)^{n_k}\ ,
\end{align}
along with the log-likelihood,
\begin{align}
  \label{eq:log-lambda-general}
  \lambda(\rho) = -2\ln\Lambda(\rho) = -2\sum_k n_k\ln\tr(E_k\rho)\ ,
\end{align}
with a conventional $(-2)$
factor~\cite{BlumeKohout2010_verification,BlumeKohout2012arXiv_Tomo}.

\heading{Confidence Regions}
In the following, we briefly review the method of Ref.~\cite{Christandl2012_Tomo} for
constructing confidence regions, on which our method is based.

Confidence regions of confidence level $1-\epsilon$ are defined as regions in state space
which contain the true state with a probability of at least $1-\epsilon$. Crucially, it is the
complete \emph{procedure} of assigning a region to tomographic data which is certified and
not the particular region itself (despite the slightly misleading terminology).  More
precisely, for a particular ``true'' state $\rho_\mathrm{true}$, the measurement outcomes
observed in the tomography procedure are only one possible outcome data set among the
enormous amount of theoretically possible data sets. Now, a data analysis procedure
associates with each observed data set a corresponding region in state space. This tomography
procedure is said to \emph{yield confidence regions of confidence level $1-\epsilon$} if,
for any true state $\rho_\mathrm{true}$, the tomography procedure associates with the
observed data set a region which contains $\rho_\mathrm{true}$, except for some data sets
with total probability $\epsilon$. In other words, the complete tomography procedure is
successful except with probability $\epsilon$, in which case the observed data set may
cause the procedure to report a bad region. These ``exceptional data sets'' may be
interpreted as misleading but highly unlikely situations. For example, if we flip a fair
coin many times and observe the sequence of all ``heads,'' any reasonable inference scheme
would wrongly report that the coin is highly biased. However this outcome only happens
with disproportionately small probability; introducing the parameter $\epsilon$ above
allows us to disregard such extremely unlikely cases.

The method of Ref.~\cite{Christandl2012_Tomo} is formulated using the \emph{estimate
  density} $\mu_{B^n}$~\footnote{We use the notation of Ref.~\cite{Christandl2012_Tomo};
  there, the general case of a joint measurement on the $n$ systems is considered and the
  outcome POVM effect is denoted by $B^n$.}, defined as
\begin{align}
  \mu_{B^n}(\rho) = \frac1{c_{B^n}}\Lambda(\rho)\ ,
  \label{eq:def-muBn}
\end{align}
where $c_{B^n}$ is a normalizing factor such that $\int d\rho\, \mu_{B^n}(\rho) = 1$, and
where $d\rho$ is the Hilbert-Schmidt measure normalized such that
$\int d\rho = 1$~\cite{BookBengtssonZyczkowski2006_Geometry,Zyczkowski2001_Induced}.  The
main result of Ref.~\cite{Christandl2012_Tomo} is a criterion for certifying a procedure
for yielding confidence regions of confidence level $1-\epsilon$.  The criterion is the
following: the procedure should map to any tomographic data (essentially) a region $R$ in
state space which satisfies
\begin{align}
  \label{eq:Gamma-region-weight-muBn}
  \int_R \mu_{B^n}(\rho)\,d\rho = 1-\frac\epsilon{\operatorname{poly}(n)}\ ,
\end{align}
i.e., which has high weight under the distribution $\mu_{B^n}$~\footnote{The output region
  should in fact be $R^\delta$, obtained by enlarging the set $R$ by some small $\delta$
  in fidelity distance; explicit expressions for $\operatorname{poly}(n)$ and $\delta$ are
  given in Ref.~\cite{Christandl2012_Tomo}.}.

\heading{Confidence Regions for a Figure of Merit}
We may now use this criterion to devise an explicit procedure for constructing confidence
regions, where the regions $R$ are chosen to be defined via level sets of a figure of
merit.

A figure of merit $f(\rho)$ may be any function of the quantum state.  For example,
$f(\rho) = F^2(\rho,\proj{\psi_\mathrm{Ref}})$ expresses the fidelity to a reference state
$\ket{\psi_\mathrm{Ref}}$.  The reduced distribution of the estimate density
$\mu_{B^n}\mathopen{}\left(\rho\right)$ onto the figure of merit $f$ is given by
\begin{align}
  \mu\mathopen{}\left(f\right)
  = \int d\rho\,\mu_{B^n}\mathopen{}\left(\rho\right)\,\delta\mathopen{}\left(f\left(\rho\right) - f\right)\ ,
\end{align}
where $\delta\mathopen{}\left(\cdot\right)$ denotes the Dirac delta function.

\begin{figure}
  \centering
  \includegraphics{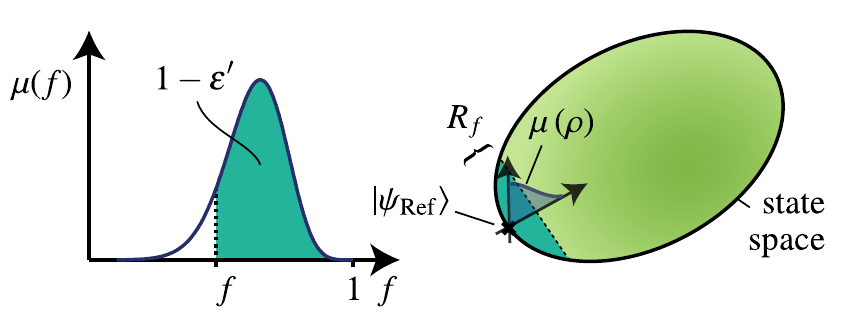}
  \caption{Construction of confidence regions from the distribution $\mu\left(f\right)$ on
    the figure of merit.  High weight intervals with respect to $\mu\left(f\right)$ (left
    plot) correspond to high weight regions in state space with respect to
    $\mu_{B^n}\left(\rho\right)$ (right diagram) which are (essentially) confidence
    regions, according to Ref.~\cite{Christandl2012_Tomo}.}
  \label{fig:ConfRegionFigureOfMeritFid}
\end{figure}

Now, fix a threshold value $f$, and consider the region $R_{f}$ in state space consisting
of all states whose figure of merit is greater than or equal to $f$
(\autoref{fig:ConfRegionFigureOfMeritFid}).  The weight of the region $R_f$ according to
the distribution $\mu_{B^n}\mathopen{}\left(\rho\right)$ is exactly given by
$\int_{f'\geqslant f}\mu\mathopen{}\left(f'\right)\,df'$.  Inverting this reasoning, for
any $\epsilon$, we can find the maximum threshold value $f$ required for a region $R_f$ to
encompass a particular weight $1-\epsilon/\operatorname{poly}\mathopen{}\left(n\right)$;
we know that this region is essentially a confidence region by the criterion of
Ref.~\cite{Christandl2012_Tomo}.  (If the figure of merit is such that smaller values of
$f\mathopen{}\left(\rho\right)$ are desirable, such as the trace distance to a reference
state, then $R_f$ is defined with $f$ as an upper, rather than lower, threshold value.)

We arrive at a first important observation: if we find a simple characterization of the
function $\mu\mathopen{}\left(f\right)$, then we are capable of constructing confidence
regions in terms of $f$ for any confidence level (See \autoref{appx:conf-reg-from-mu-f}  for
  how to transpose the $\delta$-enlargement in~\cite{Christandl2012_Tomo} into a shift of
  the threshold value $f$.).

\heading{Determining $\mu(f)$ Numerically}
We propose a practical procedure which determines a numerical estimate of
$\mu\mathopen{}\left(f\right)$.  We resort to a Monte Carlo--type scheme known as the
Metropolis-Hastings algorithm~\cite{Metropolis1953JCP} (cf.\@ also
Refs.~\cite{Shang2015NJP_sampling,Seah2015NJP_sampling2}).  This algorithm is a standard,
well-tested scheme widely used in computational physics---for instance, to simulate the
behavior of statistical systems at finite temperature~\cite{ALPS2007_1.3}---and there are
standard methods for controlling the uncertainties resulting from the use of this
procedure~\cite{Ambegaokar2010AJP_estimating}.  Using this algorithm, we conduct a random
walk in the quantum state space and produce random samples distributed according to the
distribution $\mu_{B^n}\mathopen{}\left(\rho\right)$.  By collecting the values of
$f(\rho)$ at the sampled points into a histogram, we obtain an estimate for $\mu(f)$.
(See \autoref{appx:MH-procedure} for the details of the random walk procedure).

\heading{Theoretical Model for $\mu(f)$}
It turns out that, for a selection of common figures of merit, we may understand the
numerical estimate of $\mu(f)$ with a theoretical model.  Suppose $f(\rho)$ is the
fidelity to a pure reference state, the expectation value of an observable, or the trace
distance to any reference state.  Then, under some reasonable assumptions~\footnote{%
   We assume that not too few measurements have been taken, and invoke some approximation
   methods such as Laplace's method for integrating an exponential (see details in
   \autoref{appx:fit-model-derivation}).}, we derive the
following approximate theoretical model for $\mu\mathopen{}\left(f\right)$ (see
\autoref{appx:fit-model-derivation}),
\begin{align}
  \label{eq:fit-model-p}
  \mu\left(f\right) \approx C\,\left(f-h\right)^m \cdot
  e^{-a_2\,\left(f-h\right)^2-a_1\,\left(f-h\right)}\ ,
\end{align}
with three fit parameters $a_1$, $a_2$, and $m$ (with $m\geqslant 0$), and one constant normalization
factor $C$; $h$ is a constant depending only on the choice of the figure of
merit.  Specific values of the constant $h$ for some figures of merit are summarized in
\autoref{tab:TheoModelsFiguresOfMerit}.
\begin{table}
  \centering
  \bgroup
  \def\arraystretch{1.4}
  \begin{tabular}{|>{\centering\arraybackslash}m{0.58\columnwidth}|>{\centering\arraybackslash}p{0.38\columnwidth}|}
    \hline
    \multicolumn{2}{|c|}{$\ln\mu\left(f\right)\approx -a_2 x^2 -a_1 x + m\ln x + c$,~~where:} \\
    Figure of merit $f\left(\rho\right)$ & $x=$ \\
    \hline
    $F^2\left(\rho,\proj{\psi_\mathrm{Ref}}\right)=\matrixel{\psi_\mathrm{Ref}}{\rho}{\psi_\mathrm{Ref}}$
    &  $1-f$ \\
    $D\left(\rho,\rho_\mathrm{Ref}\right) = \frac12\norm{\rho-\rho_\mathrm{Ref}}_1$
    & $f$ \\
    Observable $\langle A\rangle_\rho$ & $a-f$ or $f-a$ \\
    \hline
  \end{tabular}
  \egroup
  \caption{Theoretical fit model for some selected figures of
    merit. Here $\ket{\psi_\mathrm{Ref}}$ denotes any pure state, and $\rho_\mathrm{Ref}$ any pure
    or mixed state. We use the notation $D\left(\rho,\sigma\right)$ for the trace distance and
    $\langle A\rangle_\rho=\tr\left(A\rho\right)$ for the expectation value of an observable $A$.
    The value $a$ is an extremal value of $\langle A\rangle_\rho$ for valid density matrices $\rho$
    close to the region of interest, and $x$ should
    be chosen as $x=a-f$ (resp.\@ $x=f-a$) if $a$ is
    a maximal value (resp.\@ minimal value). If the extremum point of $A$ is far from the region of
    interest, the logarithm term in the model can be dropped, as the exponential will dominate the
    volume term, and $a$ can be absorbed into the other factors.}
  \label{tab:TheoModelsFiguresOfMerit}
\end{table}

The parameters $(a_2,a_1,m)$ are then mapped onto new parameters which are more
representative of the shape of the function.  The latter is viewed as a ``skewed''
Gaussian (see \autoref{appx:quantum-error-bars}).  The parameter $f_0$ determines the position of the
peak, the parameter $\Delta$ is the half width of the ``deskewed'' Gaussian, and
$\gamma$ characterizes the deviation from a perfect Gaussian.
The parameters $(f_0, \Delta, \gamma)$ are the \emph{quantum error bars}.

\begin{figure}
  \centering
  \includegraphics[width=76mm]{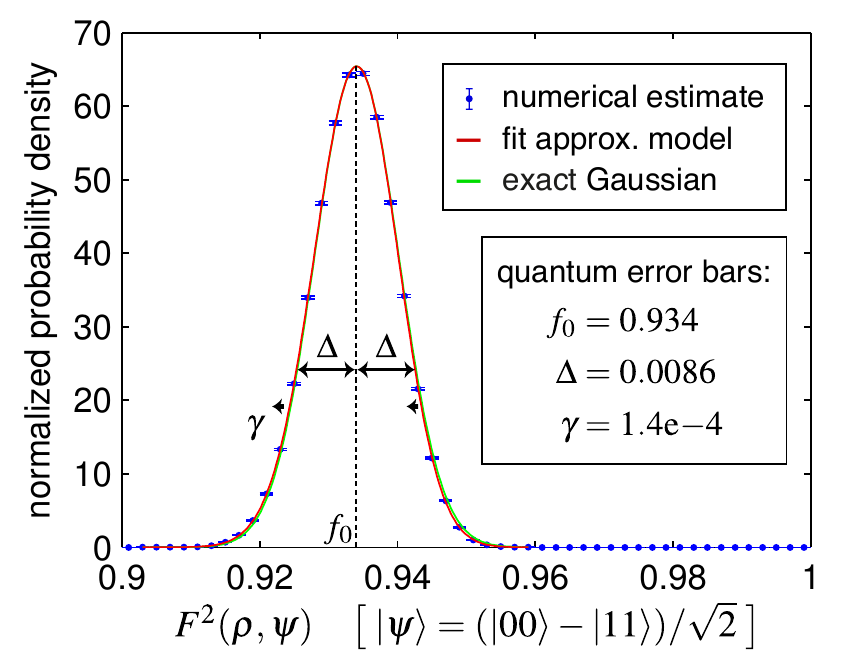}
  \caption{Analysis of measurement data from two superconducting qubits prepared in a Bell
    state. We determined effective measurement operators which model the noise in the
    measurement process. The histogram of the fidelity to the target state $\ket\psi$ (the
    blue data points), produced using our procedure, fits well to our theoretical model in
    \autoref{tab:TheoModelsFiguresOfMerit}.  The {quantum error bars} are a concise,
    intuitive, and precise characterization of the fit model, which is interpreted as a
    skewed Gaussian function.  The parameter $f_0$ is the peak maximum, $\Delta$ is the
    half width of the original Gaussian, and $\gamma$ characterizes the skewing in
    terms of the displacement of the sides of the peak from the exact Gaussian at the
    relative height $1/e$.  This example involving experimental data demonstrates a good
    level of practical applicability of our method.}
  \label{fig:AnalysisSuperconductingQubits}
\end{figure}

\heading{Application to Experimental Data}
We have applied the algorithm to experimental data from two superconducting qubits
prepared in a Bell state according to the setup described
in Refs.~\cite{Baur2012PRL_benchmarking,Steffen2013Nat_deterministic}.  The data were kindly
provided by the authors of Ref.~\cite{Steffen2013Nat_deterministic}.  The two qubits were
measured using slightly noisy individual Pauli operators, with a total of $n=55\,677$
measurements.  The numerical estimation of $\mu\left(f\right)$ corresponding to the
fidelity to the target Bell state is depicted in
\autoref{fig:AnalysisSuperconductingQubits}.  (See
\autoref{appx:application-experiment-superconducting} for details of
the analysis of the experiment, including the modeling of the
measurements~\cite{Bianchetti2009PRA} into effective
POVM operators.)

\heading{Quantum Error Bars}
The quantum error bars ($f_0$, $\Delta$, $\gamma$) displayed in
\autoref{fig:AnalysisSuperconductingQubits} are a concise and useful description
of the error analysis, from which reliable
operational statements can be made.  Indeed, they provide the necessary information for
constructing confidence regions for any given confidence level.

Also, as seen in \autoref{fig:AnalysisSuperconductingQubits}, our error bars have the
intuitive interpretation as representing the ``spread'' of the figure of merit according
to $\mu\left(f\right)$.  As such, the error bars are  much smaller than the size of a
confidence region for a small epsilon in a worst-case scenario, and they are in fact of
comparable size to those obtained by
bootstrapping~\cite{BookEfron1994Bootstrap,Mitchell2003PRL_diagnosis,Home2009Sci_iontraps,BlumeKohout2012arXiv_Tomo,Schwemmer2015PRL_systematic}
(see \autoref{appx:compare-with-bootstrap}).

\heading{Discussion}
Our work bridges the apparent gap between carrying out a mathematically rigorous,
well-justified error analysis and using an \emph{ad hoc} procedure yielding smaller error
bars.
The quantum error bars provide a convenient and precise representation of the information
provided by the tomography procedure.

While the fit model for $\mu(f)$ is subject to some assumptions and approximations, it
applies well to many examples studied by the authors in developing this work---for
$n\sim100$ total measurements already---and has been tested with up to five qubits.  Note that the
numerical procedure is not subject to these assumptions, and a deviation from the fit
model could easily be noticed in some extreme examples considered (for example, with
goodness-of-fit measures).  A further detailed discussion on the reliability of our method
is presented in \autoref{appx:reliability}.

It is relatively straightforward to apply our method to experimental setups consisting of
a few qubits.  Our procedure is restricted neither to particular measurement settings nor
to specific quantum states, and it applies, for example, to adaptive tomography.  In
general, noise in the measurement procedure has to be modeled into effective POVM effects
analogously to our approach for the two superconducting qubits. (In contrast, other
approaches do not require
this~\cite{Stark2014PRA_self,Stark2012_rigidity,Stark2012_computation}.)  We have
developed a software which implements our procedure~\cite{TomographerCxx} that is expected
to be directly applicable to most experimental settings.

For worst-case scenarios such as quantum cryptography~\cite{Tomamichel2012NC_tight}, it is
still desirable to improve the methods for explicitly constructing confidence regions.  We
do anticipate that the bounds used in Ref.~\cite{Christandl2012_Tomo} may be tightened to
yield smaller confidence regions for the same confidence level.  If the construction is
not altered, the procedure presented here would not require any change, as the same
histograms may still serve for constructing confidence regions using the tightened proof.

We also insist that our results do not rely on any particular interpretation of
``probability,'' such as a Bayesian or a frequentist one. This is because we consider experiments
which can, \emph{in principle}, be repeated arbitrarily many times, which is a regime where
these interpretations are equivalent~\cite{Christandl2012_Tomo}.  Nonetheless, the
Bayesian viewpoint is convenient, as the distribution $\mu\left(f\right)$ happens to
coincide with the Bayesian posterior corresponding to an agent starting the tomography
procedure with a Hilbert-Schmidt uniform prior.

Furthermore, even though our results are formulated in the context of quantum state
tomography, the same procedure may be applied to quantum process
tomography~\cite{Chuang1997JMO,OBrien2004PRL_controlledNOT}. Indeed, the
Choi-Jamio\l{}kowski isomorphism implies that determining a quantum process is
mathematically the same as determining a bipartite quantum state.

\heading{Acknowledgments}
We thank
 Robin Blume-Kohout,
 Matthias Christandl,
 Steve Flammia,
 Aleksejs Fomins,
 Olivier Landon-Cardinal,
 Romain M\"uller,
 Denis Rosset,
 Cyril Stark,
 Lars Steffen, and
 Takanori Sugiyama
for fruitful discussions.
We acknowledge support from the European Research Council (ERC) via Grant No.~258932, from
the Swiss National Science Foundation through the National Centre of Competence in
Research ``Quantum Science and Technology'' (QSIT), and by the European Commission via the
project ``RAQUEL.''

\clearpage

\appendix

\vspace*{2em}
\begin{center}\textbf{SUPPLEMENTAL MATERIAL}\end{center}\nopagebreak
\vspace*{3em}\nopagebreak

\makeatletter
\global\@dbltopnum\z@
\global\@dbltoproom\z@
\global\@topnum\z@
\global\@toproom\z@
\makeatother









\makeatletter
\newcounter{mysubfigcnt}[figure]
\def\mysubfigcntautorefname{Figure}
\renewcommand\themysubfigcnt{\textbf{\alph{mysubfigcnt}}}
\renewcommand\p@mysubfigcnt{\thefigure}
\setcounter{mysubfigcnt}{0}
\def\mysubfig\label#1{%
  %
  %
  \addtocounter{figure}{1}%
  \refstepcounter{mysubfigcnt}\label{#1}%
  \addtocounter{figure}{-1}%
}
\makeatother

\def\fmtcapsubfig[#1]{%
  \unskip\hspace{1.25ex plus 1ex minus 0.3ex}\textbf{#1}~\ignorespaces%
}

\makeatletter
\renewcommand*{\eqref}[1]{%
  \hyperref[{#1}]{\textup{\tagform@{\ref*{#1}}}}%
}
\makeatother

\begin{figure}[t]
  \centering
  \includegraphics[width=\columnwidth]{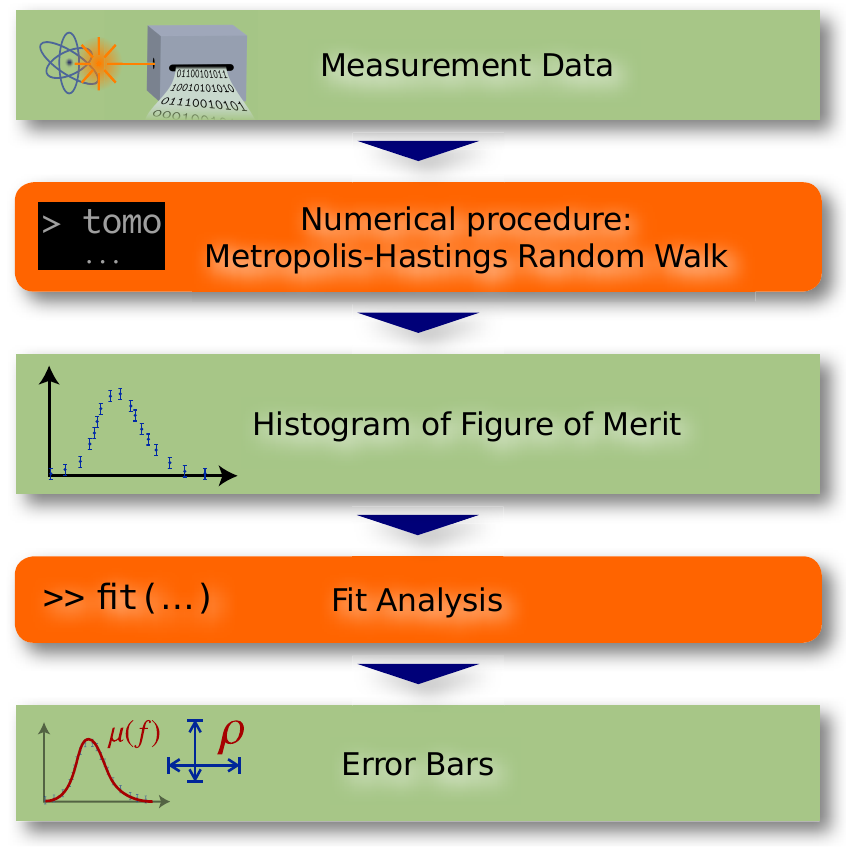}
  \caption{Overview of our quantum tomography analysis and how to apply our method. The
    measurement data are the input to our procedure. Our numerical method, which is based
    on a Metropolis-Hastings random walk in state space, outputs a histogram of a chosen
    figure of merit.  We provide software accomplishing this~\cite{TomographerCxx}.  This
    histogram is a numerical approximation to the distribution $\mu\left(f\right)$ of the
    figure of merit.  In a second step, this numerical estimate is fitted by a theoretical
    model.  The fit can be done, for instance, using MATLAB.  This yields a full
    description of the relevant distribution which allows to construct in principle
    confidence regions for any confidence level.  This description is given in terms of
    three parameters, which are then effectively the ``error bars.''}
  \label{fig:MethodSchematic}
\end{figure}

In this appendix, we provide a detailed description of how our method is
implemented, how it is applied to practical examples, as well as additional discussions
referred to from the main text.  An overview of our method is given in
\autoref{fig:MethodSchematic}.

Our software, along with instructions for download and use, may be downloaded at the location:
\url{https://tomographer.github.io/tomographer}.

\section{Procedure for determining $\mu(f)$ using a Metropolis-Hastings random walk}
\label{appx:MH-procedure}
The figure of merit is given as a function $f\left(\rho\right)$ of the quantum state.  For
example, the squared fidelity to a reference state $\rho_\mathrm{Ref}$ is represented as
$f\left(\rho\right)=F^2\left(\rho,\rho_\mathrm{Ref}\right)$.

Recall that the relevant object in the method of Christandl and Renner is the
\emph{estimate density} $\mu_{B^n}\left(\rho\right)$, given by
Eq.~(\ref*{eq:def-muBn}) of the main text.

Given the figure of merit $f\left(\rho\right)$ of interest, the reduced distribution on this
of merit of $\mu_{B^n}\left(\rho\right)$ is
\begin{align}
  \label{eq:def-mu-of-f}
  \mu\left(f\right)
  = \int d\rho\,\mu_{B^n}\left(\rho\right)\,\delta\left[f\left(\rho\right) - f\right]\ ,
\end{align}
where the Dirac delta ensures the integration is performed over the shell of states in
state space which have the given figure of merit $f$. The quantity $\mu\left(f\right)$
corresponds to the total weight given by $\mu_{B^n}\left(\rho\right)$ to all states with a
given fixed figure of merit $f$.

In the following, we develop a method to compute $\mu\left(f\right)$ numerically.  We
resort to a Monte Carlo-type scheme known as the Metropolis-Hastings
algorithm~\cite{Metropolis1953JCP} (cf.\@ also
Refs.~\cite{Shang2015NJP_sampling,Seah2015NJP_sampling2}). This scheme is widely used in
computational physics, for instance to simulate the behavior of statistical systems at
finite temperature~\cite{ALPS2007_1.3}. This algorithm conducts a random walk which
produces random samples distributed according to a given distribution $P\left(x\right)$.
The parameters of the algorithm are a starting point $x_0$ as as well as a ``jump
distribution'' $Q\left(x'|x\right)$. The jump distribution is assumed to be symmetric
($Q\left(x'|x\right)=Q\left(x|x'\right)$), and is used to update the current step in the
random walk.  (For example, $Q\left(x'|x\right)$ is often chosen as a Gaussian in some
relevant coordinates centered at $x$).  The $i$-th step of the random walk goes as
follows:
\begin{enumerate}[label=\arabic*.]
\item Choose a new candidate point $x'$ according to $Q\left(x'|x_i\right)$;
\item Calculate $a = P(x') / P(x_i)$. If $a>1$, then set $x_{i+1}:=x'$ unconditionally; if
  $a<1$, then decide randomly to set $x_{i+1}:=x'$ with probability $a$, or else to set
  $x_{i+1}:=x_i$.
\end{enumerate}
The sequence of points $\{x_i\}$, albeit correlated, are then asymptotically distributed
according to the distribution $P\left(x\right)$.

In order to calculate the quantity $\mu\left(f\right)$, we draw a large number of random
samples in the quantum state space according to the distribution
$\mu_{B^n}\left(\rho\right)$, i.e.\@ with $\mu_{B^n}\left(\rho\right)$ playing the role of
$P\left(x\right)$. The histogram of values $f\left(\rho\right)$ evaluated at those samples
then provide a numerical estimate of $\mu\left(f\right)$. Crucially, it is not necessary
to calculate the normalization constant $c_{B^n}$ in the definition of $\mu_{B^n}$
(Eq.~(\ref*{eq:def-muBn}) of the main text) because in the Metropolis-Hastings
algorithm we only have to evaluate ratios of probabilities.

For our random walk, we represent a quantum state $\rho$ of dimension $d$ by a square
complex matrix $T$ with $\tr TT^\dagger=1$, such that $\rho = TT^\dagger$. To any such $T$
corresponds a valid density matrix $\rho$, and to any density matrix $\rho$ corresponds at
least one such $T$ (e.g.\@ $T=\rho^{1/2}$). Additionally, the constraint
$\tr TT^\dagger=1$ corresponds to requiring that the components of $T$, real and imaginary
parts taken separately into a real vector $\vec y$, lie on the surface of the unit
$(2d^2-1)$-hypersphere. Random density matrices may be sampled from the Hilbert-Schmidt
measure by choosing such random points on this
hypersphere~\cite{BookBengtssonZyczkowski2006_Geometry}. In fact, the matrix entries
$T_{ij}$ of $T$ are simply the components of a vector $\ket\psi$ of dimension $d^2$ which
purifies $\rho$. Indeed, if we trace out the second system from
$\ket\psi = \sum_{ij} T_{ij} \ket i_A\otimes\ket j_B$, we obtain
$\tr_B\proj\psi = \sum_{iji'} T_{ij}T_{i'j}^*\ketbra{i}{i'} = TT^\dagger = \rho$. We hence
choose to perform a Metropolis-Hastings random walk on the $(2d^2-1)$-hypersphere
corresponding to the possible $T$ matrices. The jump candidate is calculated from a point
$\vec y_i$ by choosing a vector $\vec\omega$ of $2d^2$ normally distributed values and
setting
$\vec y' = (\vec y_i + \eta_\mathrm{step}\vec\omega)/\norm{\vec
  y_i+\eta_\mathrm{step}\vec\omega}$,
where $\eta_\mathrm{step}$ is a chosen step size. The jump distribution obtained in this
way is symmetric.

We follow the the prescriptions given in Ref.~\cite{Ambegaokar2010AJP_estimating} for the
correct usage and appropriate error analysis of the Metropolis-Hastings algorithm.
First, since the random starting point is likely to be a point which has very little
weight under $P\left(x\right)$, the random walk needs to equilibrate, or thermalize, until
it reaches points which have a non-negligible values of $P\left(x\right)$. This first set
of points traversed until the walk has thermalized should be discarded.
Also, because consecutively collected samples may be very correlated, it is useful to keep
only one sample in a certain number $N_\mathrm{sweep}$ (the ``sweep size''), while
throwing away each time the $N_\mathrm{sweep}-1$ points between two recorded samples. In
our examples, the sweep size $N_\mathrm{sweep}$ is chosen of the order of
$1/\eta_\mathrm{step}$; this gives at least the chance to the random walk to traverse all
of state space between two recorded samples.
Errors on the final numerical histogram points may be estimated either by calculating the
standard deviation of independent runs of the simulation, or by a binning analysis which
takes into account the correlation of the samples during a single run. We refer to
Ref.~\cite{Ambegaokar2010AJP_estimating} for a detailed discussion of these
techniques.

In our case, for each histogram bin, we associate to each recorded sample the value~$1$ if
the point is in the bin, or~$0$ otherwise. The final numerical estimate of
$\mu\left(f\right)$ is produced by averaging the time series for each bin, which
corresponds up to a constant to generating a histogram of counts. These time series are
suitable for use in a binning analysis to obtain error bars on the numerical estimate of
$\mu\left(f\right)$.

\section{The fit model for $\mu\left(f\right)$}
\label{appx:fit-model-derivation}
We now derive an approximate theoretical model to fit our numerical histogram. This is
useful in order to provide a succint description of the result in terms of only a few
parameters. It also serves as a consistency check allowing us to assert that our results
are well understood from a theoretical point of view. 

In general, the function $\mu\left(f\right)$ can be very complicated, so an exact
analytical description is unlikely. Rather, our goal is to find a decent approximation of
this function in a region close to where $\mu_{B^n}$ has high weight.

In fact, the bell shape of the curves in
\autoref{fig:AnalysisSuperconductingQubits} of the main text is typical when the
figure of merit is taken to be the fidelity to a pure target state, the expectation value
of an observable or the trace distance to any reference state.  Intuitively, this shape is
the result of
the concurrence of two effects: a volume factor reflecting the increasing surface of a
shell of fixed figure of merit as we get far from the reference point, and the
approximately exponential decrease of the likelihood function itself. For example, in the
case of the trace distance to the maximum likelihood estimate, there are many more states
with high distance to $\hat\rho_\mathrm{MLE}$ than there are very close---this is the
increasing volume factor.  The following derivation makes this argument more precise.

Let's now derive the fit model.  We parameterize $\rho$ with a generalized Bloch
vector~\cite{Bertlmann2008JPA_qudits,Bruning2012_parametrizations}. Take an orthonormal
basis $\{A_j\}$ of the Lie algebra $su(d)$, with $j=1,\ldots,M$ and $M=d^2-1$. The $A_j$
are hermitian, traceless and obey $\tr A_j A_{j'}=\delta_{jj'}$ (an example are the
normalized generalized Gell-Mann
matrices~\cite{Bertlmann2008JPA_qudits,Bruning2012_parametrizations}, or, for $k$ qubits,
the normalized tensor product of Pauli operators). We may now write a general state $\rho$
as $\rho\left(a_j\right)=(1/d)\Ident + \sum_j a_j A_j$ with real coefficients $a_j$
obeying some nontrivial constraints such that $\rho\geqslant0$. The Hilbert-Schmidt
distance is given by the Euclidean distance of the generalized Bloch vectors,
$\tr\bigl[\bigl(\rho(a_j)-\rho(b_j)\bigr)^2\bigr] = \sum_j \bigl(a_j-b_j\bigr)^2$. Now
because the Hilbert-Schmidt measure $d\rho$ is induced by the Hilbert-Schmidt
metric~\cite{Zyczkowski2001_Induced,BookBengtssonZyczkowski2006_Geometry}, we may
write~\eqref{eq:def-mu-of-f} as
\begin{align}
  \mu\left(f\right) = \frac1{c'}\,\int d^M a_j\, e^{-\frac12\lambda(a_j)}\,
  \delta\left[f(a_j) - f\right]\ ,
\end{align}
with a new constant $c'$ and $\lambda\left(\rho\right) = -2\ln\Lambda\left(\rho\right)$,
and where implicitly the arguments to $\lambda\left(\cdot\right)$ and
$f\left(\cdot\right)$ are to be transformed into $\rho$ appropriately.

The conditions that the $a_j$ have to satisfy in order to represent a positive
semidefinite matrix are
complicated~\cite{Bertlmann2008JPA_qudits,Kimura2003PLA_Bloch,Byrd2003PRA_positivity}.
However, it turns out that we don't need to know the exact form of these constraints.
Rather, we assume that:
\begin{enumerate}[label=(\roman*)]
\item $f$ has an extremal value close to the region of interest (\emph{viz.}, near
  $\rho_\mathrm{MLE}$);
\item the surface of a shell of states of a given figure of merit $f$ tends to zero as $f$
  tends to this extremal value.
\end{enumerate}
These assumptions are rather natural and are indeed automatically satisfied if
$f\left(\rho\right)$ is one of the cases considered in the main text (the squared fidelity
to a pure reference state, the trace distance to a reference state $\rho_\mathrm{Ref}$, or
the expectation value of an observable).  In the case of a distance measure, such as the
trace distance, the extremum is usually zero
at the reference state itself, and the surface of the shell of
states with very small distance to $\rho_\mathrm{Ref}$ clearly shrinks to zero. In the
case of the expectation value of an observable, the extremum is attained at the border of
state space. Because the border of state space has no flat facets, the surface of a shell
of given expectation value (a hyperplane intersected with state space) also shrinks to
zero as we approach the border.
Furthermore recall that the squared fidelity to a pure reference state
$\ket{\psi_\mathrm{Ref}}$ can be written as the expectation value of the observable
$\proj{\psi_{\mathrm{Ref}}}$.

Denote by $\rho_\mathrm{Ref}$ a relevant reference point where the figure of merit is
extremal, and let $a^\mathrm{Ref}_j$ such that $\rho_\mathrm{Ref} = (1/d)\Ident + \sum_j
a^\mathrm{Ref}_j A_j$.  (We recycle the notation $\rho_\mathrm{Ref}$ since whenever the
figure of merit is a distance measure to a reference state, the same reference state is to
be used here.)  We go to hyperspherical coordinates $(r,\Omega)$ centered at the reference
point $a^\mathrm{Ref}_j$, with $d^M a_j = dr\,d\Omega\,r^{M-1}$, and introduce the change
of variables $r\to r' = f\left(r,\Omega\right)$:
\begin{align}
  \mu\left(f\right) &= \frac1{c'}\,\int dr\,d\Omega\,r^{M-1}\,
  e^{-\frac12\lambda\left(r,\Omega\right)} \,
  \delta\left[f\left(r,\Omega\right) - f\right]
  \nonumber\\
  &= \frac1{c'}\,\int dr'\,d\Omega
  \left[r^{M-1}\,\abs*{\frac{\partial f}{\partial r}}^{-1}\right]
  e^{-\frac12\lambda(r',\Omega)} \,
  \delta\left[r' - f\right]
  \nonumber\\
  &= \frac1{c'}\,\int d\Omega
  \left[ r^{M-1}\,\abs*{\frac{\partial f}{\partial r}}^{-1}\right]
  e^{-\frac12\lambda\left(f,\Omega\right)}\ ,
    \label{eq:calc-fit-model-intermediate-exact}
\end{align}
where in the last two integrals the terms in square brackets are to be evaluated at the
points $r$ which satisfy $r' = f(r,\Omega)$ and $f=f(r,\Omega)$, respectively. Note that
the figure of merit $f(r,\Omega)$ must be invertible for fixed $\Omega$ and for
$r\geqslant 0$; this is usually the case with our choice of $a^\mathrm{Ref}_j$
above. Note also that Expression~\eqref{eq:calc-fit-model-intermediate-exact} is in fact
still exact, albeit very difficult to calculate explicitly. To proceed further, we will use
Laplace's approximation, and assume that the main contribution to the integral is a region
close to a single point $\Omega_0$ on the shell of fixed figure of merit $f$ where the
integrand is maximal. Then, we have
\begin{align}
  \text{\eqref{eq:calc-fit-model-intermediate-exact}} \approx
  \frac1{c'}\left[ r^{M-1}\,\abs*{\frac{\partial f}{\partial r}}^{-1} \right]_{\Omega_0}
  w\left[f,\Omega_0\right]\, e^{-\frac12\lambda\left(f,\Omega_0\right)}\ ,
  \label{eq:calc-fit-model-after-Laplace}
\end{align}
where $w\left[f,\Omega_0\right]$ is a ``width factor,'' which accounts for the total
weight of the peak in Laplace's approximation, including a possible truncation of the peak
caused by the border of state space.  Note that the condition (ii) above for the figure of
merit ensures that $w\left[f,\Omega_0\right]$ doesn't explode when $f$ approaches the
extremum of $f\left(\rho\right)$.

At this point, we need to make further assumptions about the behavior in $f$ of the
individual terms in~\eqref{eq:calc-fit-model-after-Laplace}. First, we consider the regime
in which the likelihood is close to Gaussian in the Hilbert-Schmidt coordinates,
\begin{align}
  \label{eq:lambda-2ndorder-Gaussian-approx}
  \lambda\left(\vec a\right) \approx \lambda_0 + \vec \lambda_A\cdot(\vec a-\vec
  a^\mathrm{Ref}) + (\vec a-\vec a^\mathrm{Ref})^T \lambda_B\,(\vec a - \vec
  a^{\mathrm{Ref}}),
\end{align}
(we shift $\vec a$ by $\vec a^\mathrm{Ref}$ without loss of generality and for later
practical reasons; also $\lambda_B$ is a symmetric matrix).  This is generically the case
for most practical scenarios where a reasonable amount of measurements were taken.

Second, we need to assume something about the figure of merit $f\left(r,\Omega\right)$:
we'll suppose that
\begin{align}
  \label{eq:assumption-property-f-affine-in-r}
  f\left(r,\Omega\right) = r\,g\left(\Omega\right)+h\ ,
\end{align}
where $h$ is some known constant, and $g\left(\Omega\right)$ some function.  The figures
of merit considered in the main text automatically satisfy this assumption.  First,
if the figure of merit is any distance measure to $\rho_\mathrm{Ref}$ which is given by a
norm, such as the trace distance, then
$f\left(r,\Omega\right) = \norm{\rho\left(r,\Omega\right) - \rho_\mathrm{Ref}} =
\norm{\sum_j (a_j(r,\Omega) - a^\mathrm{Ref}_j)\; A_j } = r\norm{\sum_j\Omega_jA_j}
$, recalling that our hyperspherical coordinates are defined by $\vec a(r,\Omega) = \vec
a^\mathrm{Ref} + r\,\vec\Omega$ with $\vec\Omega$ the unit vector in the direction
$\Omega$. Also, $f\left(\rho\right)$ obeys
property~\eqref{eq:assumption-property-f-affine-in-r} if it is the expectation value of an
observable, $f\left(\rho\right)=\tr\left(\rho W\right)$: with $\vec a =
r\,\vec\Omega + \vec a^\mathrm{Ref}$ we can write $f\left(\rho\right) =
\tr\left(\left[(\Ident/d)+\sum_j a_j A_j\right]W\right) = r\:\vec\Omega\cdot\vec w + (\vec
a^\mathrm{Ref}\cdot\vec w + \tr W/d)$, where $\vec w$ is the vector with components
$w_j=\tr\left(A_j W\right)$. Recall in this case that $\vec a^\mathrm{Ref}$ is on the
border of state space.   Furthermore, recall that the squared fidelity to a pure reference
state can be written as the expectation value of an observable,
$F^2(\rho,\proj{\psi_\mathrm{Ref}})=\tr(\rho\proj{\psi_\mathrm{Ref}})$.
However, if the figure of merit is the fidelity or purified distance to a mixed
reference state, it does not satisfy in general the
Ansatz~\eqref{eq:assumption-property-f-affine-in-r}.

Armed with both assumptions~\eqref{eq:lambda-2ndorder-Gaussian-approx}
and~\eqref{eq:assumption-property-f-affine-in-r}, we see that $\partial f/\partial r =
g(\Omega)$ as well as $r = \left(f - h\right)/g\left(\Omega\right)$, and we obtain
\begin{multline}
  \label{eq:fit-model-p-calc-all-terms}
  \mu\left(f\right) \approx \frac1{c''}
  \left[ \frac{\left(f - h\right)^{M-1}}{g^{M}\left(\Omega_0\right)}
  \right] \cdot w\left[f,\Omega_0\right]
  \\
  \cdot
  e^{-\frac12\left[\lambda_0 + r\vec\lambda_A\cdot\vec\Omega_0
      + r^2\vec\Omega_0^T\lambda_B\vec\Omega_0\right]}\ ,
\end{multline}
where $c''=c'\cdot\operatorname{sign}\left(g\left(\Omega_0\right)\right)$.  The term in
the exponential in~\eqref{eq:fit-model-p-calc-all-terms}, being quadratic in $r$, is then
also quadratic in $f-h$. At this point, we further assume that $\Omega_0$ (where
$\lambda\left(f,\Omega_0\right)$ is minimal at fixed $f$) is approximately constant in
$f$, and that the term $w\left[f,\Omega_0\right]$ is either approximately constant or,
being a volume factor, varies as a power of $r$, and thus of $f-h$. We finally obtain our
fit model,
\begin{align}
  \mu\left(f\right) \approx C\,\left(f-h\right)^m \cdot
  e^{-a_2\,\left(f-h\right)^2-a_1\,\left(f-h\right)}\ ,
\end{align}
with 3 fit parameters $a_1$, $a_2$, $m$ and one constant normalization factor $C$. The
value $m$ includes the $(M-1)$ power plus any contribution from the weight factor
$w\left[f,\Omega\right]$.  The expression for the logarithm of $\mu\left(f\right)$ is
numerically more suitable for fitting. We thus obtain our fit model for
$\ln\mu\left(f\right)$,
\begin{widetext}
\begin{subequations}
  \label{eq:fit-model-logp}
  \begin{align}
    \raisebox{-2ex}[0pt][0pt]{$\ln \mu\left(f\right)\ \approx\ \Bigg\{$}\hspace*{0.3ex}
    & - a_2\left(f-h\right)^2 - a_1\left(f-h\right)
    + m\,\ln\left(f-h\right) + c\ ,\text{~or} \label{eq:fit-model-logp-f-h}\\
    & - a_2\left(h-f\right)^2 - a_1\left(h-f\right)
    + m\,\ln\left(h-f\right) + c\ , \label{eq:fit-model-logp-h-f}
  \end{align}
\end{subequations}
\end{widetext}
depending on whether $f\geqslant h$ for all valid $f$ or $f\leqslant h$ for all valid
$f$. Indeed, either of these two conditions hold as we have chosen the center
$a^\mathrm{Ref}_j$ of our hyperspherical coordinates as an extremal point of
$f\left(\rho\right)$. \autoref{tab:TheoModelsFiguresOfMerit} of the main text
summarizes the appropriate fit model for a selection of figure of merits.

It is further worth mentioning that for larger $f$, the exponential will dominate all the
other terms; for example, in this regime, the details of the function
$w\left(f,\Omega\right)$ is not relevant for most figures of merit.

There are certain situations in which our approximate fit model fails to accurately
describe the behavior of $\mu\left(f\right)$.  If too few measurements are taken, the
likelihood function is not approximately Gaussian as we have assumed (however this is
usually the case already for, e.g., $n\sim100$ total measurements). Our derivation also no
longer applies if $\Omega_0$ happens to not be constant with $f$, or if a different figure
of merit is considered such as the fidelity to a mixed reference state. Furthermore, in
some cases the boundary of state space might interfere with our approximation (it might
for example constrain $\Omega_0$ causing it to vary with $f$), or the Laplace method might
not be a good approximation if $\lambda\left(f,\Omega\right)$ has e.g.\@ several minima
for fixed $f$.  However in examples we have studied these cases always caused our model to
fit poorly to the numerical estimate in the region of the peak; we consider it very
unlikely that the fit model fits the peak well but fails to describe the tail accurately.
See also \autoref{appx:reliability} for a more general discussion of the reliability of our
method.

\section{The quantum error bars}
\label{appx:quantum-error-bars}
\begin{figure}
  \centering
  \mysubfig\label{fig:ModelFunctionSkewing}
  \mysubfig\label{fig:QuantumErrorBarsSignificance-x}
  \mysubfig\label{fig:QuantumErrorBarsSignificance-f}
  \includegraphics[width=\columnwidth]{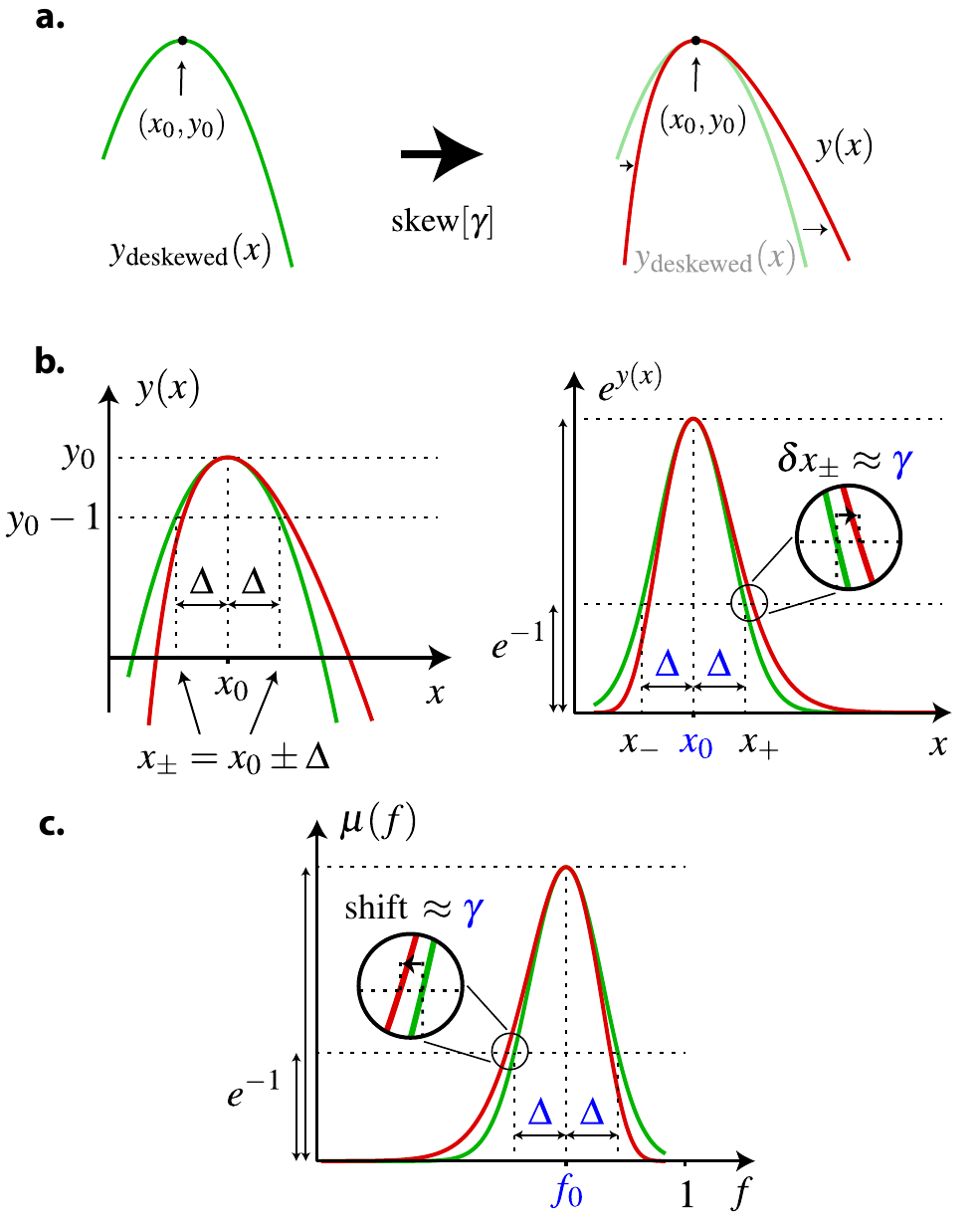}
  \caption{Model function for $\mu(f)$ as skewed Gaussian. \fmtcapsubfig[a.] The function
    $y(x)=-a_2\,x^2-a_1\,x+m\,\ln x+c$ (used to model $\ln\mu(f)$) can be seen as a skewed
    parabola, with a ``skewing operation'' parametrized by $\gamma$.  The green curve
    $y_\mathrm{deskewed}(x)$ is an actual parabola.  \fmtcapsubfig[b.] The model function
    $y(x)$ is fully characterized by the location of the peak $x_0$, the half width
    $\Delta$ of the ``de-skewed'' parabola at relative height $1/e$, and a parameter
    $\gamma$ characterizing the shift of the sides of the peak at relative height $1/e$.
    \fmtcapsubfig[c.] The variable $f$ relates to $x$ via a simple shift and possible
    reflection, given by $x=s\,(f-h)$ for a constant $h$ and $s=\pm1$ depending on the
    figure of merit (see \autoref{tab:TheoModelsFiguresOfMerit} of the main
    text).  Hence, with $f_0=s\,x_0+h$, the curve $\mu(f)$ is fully characterized by the
    parameters $(f_0,\Delta,\gamma)$, which we call the \emph{quantum error bars}.}
  \label{fig:ModelFunctionSkewing_composed}
\end{figure}

Now we proceed to transform the parameters $(a_2, a_1, m)$ into more meaningful
quantities, corresponding to distinctive features of the corresponding function.  Consider
the function
\begin{align}
  y(x) = -a_2\, x^2 - a_1\, x + m\,\ln(x) + c\ ,
  \label{eq:model-function-y-of-x-given-by-a2-a1-m-c}
\end{align}
which for $x>0$ exhibits a characteristic skewed bell curve as depicted in
\autoref{fig:AnalysisSuperconductingQubits} of the main text.  The function
$y(x)$ is exactly the model for $\ln\mu(f)$, with $x=s\,(f-h)$ for a constant $h$ and for
a sign $s=\pm1$ depending on the figure of merit, as given by~\eqref{eq:fit-model-logp}.
The function~\eqref{eq:model-function-y-of-x-given-by-a2-a1-m-c} can be seen as a skewed
version of a parabola with summit at $(x_0, y(x_0))$ (see
\autoref{fig:ModelFunctionSkewing}).
The de-skewing operation applied to $y(x)$ consists in finding the parabola
$y_\mathrm{deskewed}(x)=-a\,(x-x_0)^2+y_0$ with the same the peak position and curvature
as $y(x)$.  In other words, we find a parabola $y_\mathrm{deskewed}(x)$ such that the
functions $y(x)$ and $y_\mathrm{deskewed}(x)$ must match at $x=x_0$ to second order.
The maximum of $y(x)$ is at the point $x_0$ satisfying $0=\left.(d y/d x)\right|_{x_0}$,
that is, $0=-2\,a_2 x_0 - a_1 + m/x_0$.  Solving for $x_0$ while ensuring that $x_0>0$
yields
\begin{align}
  x_0 = \frac1{4\,a_2}\left( - a_1 + \sqrt{a_1^2 + 8\,a_2\,m} \right)\ .
\end{align}
  The condition
$(d^2y/dx^2)|_{x_0}=(d^2y_\mathrm{deskewed}/dx^2)|_{x_0}$ yields
$ a = a_2 + m/(2\,x_0^2) $.
In terms of $(a,x_0,y_0)$, the original parameters $(a_2,a_1,c)$ read
\begin{subequations}
  \begin{align}
    a_2 &= a - \frac{m}{2\,x_0^2}\ ; \\
    a_1 &= \frac{2m}{x_0}-2\,a\,x_0\ ; \\
    c &= y_0 + a_2\, x_0^2 + a_1\, x_0 - m\ln x_0\ .
  \end{align}
\end{subequations}
We can already define $\Delta$, which is the first of our quantum error bars.  It is
defined as
\begin{align}
  \Delta = \frac1{\sqrt a} = \left(a_2 + \frac{m}{2\,x_0^2}\right)^{-1/2}\ .
\end{align}
The parameter $\Delta$ is the half width of the Gaussian function
$e^{y_\mathrm{deskewed}(x)}$ at relative height $1/e$
(\autoref{fig:QuantumErrorBarsSignificance-x}): Indeed, the standard deviation of a Gaussian
is precisely the half width of the Gaussian peak at relative height $1/e$ with respect to
the Gaussian peak maximum.  In our case, $\Delta$ is interpreted as the half width of the
Gaussian, before the skewing operation is applied.

It remains to understand the effect of the $m$ parameter in terms of skewing.  Consider
the intercepts of $y_\mathrm{deskewed}(x)$ with the line $y=y_0-\xi$, which are at
$x_\pm=x_0\pm\xi^{1/2}\Delta$.  (These points correspond to the cross-section of the peak
of $e^{y_\text{deskewed}(x)}$ at a relative height $e^{-\xi}$.)  If we view the function
$y(x)$ as the result of skewing $y_\mathrm{deskewed}(x)$ via the transformation above
parametrized by $m$, then the intercepts with the line $y=y_0-\xi$ are shifted by some
$\delta x_\pm$ which vary as a function of $m$.  Let us determine $\delta x_\pm$ to first
order in $m$.  For infinitesimal $m$, the equation $y(x_\pm)=y_0-\xi$ defining $x_\pm$
varies correspondingly as $y(x_\pm+\delta x_\pm)+\delta y(x_\pm + \delta x_\pm)=y_0-\xi$.
Keeping only the terms of first order in $m$ we obtain
\begin{align}
  \left.\frac{dy}{dx}\right|_{x_\pm}\delta x_\pm + \delta y(x_\pm) = 0\ .
  \label{eq:quantum-error-bars-variation-of-xpm-under-m-condition}
\end{align}
Noting that we only need $(dy/dx)|_{x_\pm}$ to zeroth order in $m$, we have
\begin{multline}
  \left.\frac{dy}{dx}\right|_{x_\pm,m=0} = -2\,a_2\,x_\pm - a_1
  = -2\,a\,x_\pm + 2\,a\,x_0
  \\ = \mp 2\,a\,\xi^{1/2}\Delta\ .
\end{multline}
Also, with $\delta a_2 = -m/(2x_0^2)$, $\delta a_1=2m/x_0$ and $\delta c = (\delta a_2)
\,x_0^2 + \delta a_1 - m\ln x_0$, we have
\begin{widetext}
\begin{align}
  \delta y(x_\pm)
  &= -(\delta a_2)\,x_\pm^2-(\delta a_1)\,x_\pm + m\,\ln(x_\pm) + \delta c
    \nonumber\\
  &= (\delta a_2)(x_0^2-x_\pm^2) + (\delta a_1)(x_0-x_\pm) + m\,\ln\frac{x_\pm}{x_0}
    \nonumber\\
  &= -\frac{m}{2x_0^2}\cdot\bigl(\mp 2x_0\xi^{1/2}\Delta - (\xi^{1/2}\Delta)^2\bigr)
    \mp \frac{2m}{x_0}\cdot(\xi^{1/2}\Delta) + m\,\ln\left(1 + \frac{\xi^{1/2}\Delta}{x_0}\right)
    \nonumber\\
  &= \frac{\mp m\xi^{1/2}\Delta}{x_0} + \frac{m\,(\xi^{1/2}\Delta)^2}{2\,x_0^2} +
    m\,\ln\biggl(1\pm\frac{\xi^{1/2}\Delta}{x_0}\biggr)\ .
  \label{eq:quantum-error-bars-variation-calc-1}
\end{align}
Developing the logarithm as a Taylor series in $\Delta$, the first two orders cancel and
we have
\begin{align}
  \text{\eqref{eq:quantum-error-bars-variation-calc-1}}
  &\approx \frac{\pm m\,(\xi^{1/2}\Delta)^3}{3\,x_0^3}
  - \frac{m\,(\xi^{1/2}\Delta)^4}{4\,x_0^4} + \cdots
  + \frac{(-1)(\mp1)^k\xi^{k/2}\Delta^{k}}{k\cdot x_0^k} + \cdots
    \ .
\end{align}
Then we obtain from~\eqref{eq:quantum-error-bars-variation-of-xpm-under-m-condition},
also recalling that $a=\Delta^{-2}$,
\begin{align}
  \delta x_\pm
  &= -\Bigl(\frac{dy}{dx}\Bigr|_{x_\pm}\Bigr)^{-1}\cdot\delta y|_{x_\pm} \nonumber\\
  &= -\bigl(\mp 2\,\xi^{1/2}\Delta^{-1}\bigr)^{-1}
    \left(\frac{\pm m(\xi^{1/2}\Delta)^3}{3\,x_0^3} + \cdots
    + \frac{(-1)(\mp1)^k\xi^{k/2}\Delta^{k}}{k\cdot x_0^k} + \cdots \right)
    \nonumber\\
  &= \frac{m\xi\Delta^4}{6\,x_0^3} \mp \frac{m\xi^{3/2}\Delta^5}{8\,x_0^4} + \cdots
  + \frac{(\mp1)^{k+1}\,\xi^{(k-1)/2}\,\Delta^{k+1}}{2\,k\cdot x_0^{k}} + \cdots
  \ .
\end{align}
\end{widetext}
We now introduce the \emph{skewing factor} $\gamma$ as
\begin{align}
  \gamma=\frac{m\,\Delta^4}{6\,x_0^3}\ ,
\end{align}
such that to lowest order in $\Delta$, the shift of the ``sides of the peak'' given by
$x_\pm$ for a relative height $e^{-\xi}$ is directly proportional to $\gamma$
(see also \autoref{fig:QuantumErrorBarsSignificance-x}):
\begin{align}
  \delta x_\pm \approx \xi\,\gamma\ .
\end{align}
More precisely, the shift for infinitesimal $m$ is given by
\begin{multline}
  \delta x_\pm = \gamma\cdot\bigg(\xi \mp \frac{3\,\xi^{3/2}\Delta}{4\,x_0} + \cdots +
  \\
    \frac{(\mp1)^{k'}\cdot3\cdot\xi^{(k'/2)+1}\Delta^{k'}}{(k'+3)\cdot x_0^{k'}} + \cdots\bigg)\ .
\end{multline}

We may straightforwardly define $f_0=s\,x_0+h$ as the position of the peak in terms of the
figure of merit $f$, by invoking the relation $x=s\,(f-h)$ which we used to
write~\eqref{eq:model-function-y-of-x-given-by-a2-a1-m-c}.  Finally, we obtain a set of
parameters $(f_0, \Delta, \gamma)$, along with a normalization constant $y_0$, which now
all have a direct interpretation in terms of features of the modeled distribution
(\autoref{fig:QuantumErrorBarsSignificance-f}).  In summary, they are given in terms of
the fitted parameters $(a_2,a_1,m)$ as:
\begin{subequations}
  \label{eq:QuantErrorBars}
  \begin{align}
    f_0 &= h+\frac{s}{4\,a_2}\left( - a_1 + \sqrt{a_1^2 + 8\,a_2\,m} \right)\ ;
          \label{eq:QuantErrorBar-f0} \\
    \Delta &= \left(a_2 + \frac{m}{2\,x_0^2}\right)^{-1/2}\ ;
             \label{eq:QuantErrorBar-Delta} \\
    \gamma &= m\cdot\frac{\Delta^4}{6\,x_0^3}\ ,
             \label{eq:QuantErrorBar-gamma}
  \end{align}
\end{subequations}
recalling that $s=\pm1$ and $h$ in the relation $x=s\,(f-h)$ are fixed by the choice of
figure of merit, as given in \autoref{tab:TheoModelsFiguresOfMerit} of the main
text.
The position of the peak is at $f=f_0$.  The half width of the peak (after it is
de-skewed) is given as $\Delta$ (at relative height $1/e$), and the factor $\gamma$
measures how much the peak is skewed towards larger $f$ values (respectively lesser $f$
values, if $s=-1$), with a direct interpretation in terms of the horizontal shift of the
sides of the peak.

\section{Confidence regions from the distribution $\mu\left(f\right)$}
\label{appx:conf-reg-from-mu-f}

Here we see how to construct confidence regions from the histogram obtained by our method.
As explained in the main text, regions with high weight in state space may be promoted to
confidence regions by the method of Christandl and Renner.

Consider the region with all states $\rho$ which have at least a given value of the figure
of merit:
\begin{align}
  \label{eq:def-R-f-from-histogram}
  R_f = \left\{ \rho: f\left(\rho\right) \geqslant f \right\}\ .
\end{align}

The direction of the inequality in~\eqref{eq:def-R-f-from-histogram} depends on which
figures of merit are considered desirable. The direction used here reflects the fidelity
to a target state, in which case higher fidelities are desirable. If, e.g., a proper
distance measure such as the trace distance is used, the opposite inequality is
preferable.

It is straightforward to see that the weight of the region $R_f$ in state space according
to the measure $\mu_{B^n}\left(\rho\right)d\rho$ is directly given by the weight of the
function $\mu\left(f\right)$ over the corresponding range of $f$ values which are included
in the region $R_f$. For example, if the figure of merit is the fidelity to a target
state, the weight $\alpha\left(f\right)$ of the region $R_f$ is given by
\begin{align}
  \label{eq:weight-of-region-Rf}
  \alpha\left(f\right) = \int_{\rho\in R_f} d\rho\,\mu_{B^n}\left(\rho\right)
  = \int_f^1 df'\mu\left(f'\right)\ .
\end{align}

The value of $f$ required for a region $R_f$ to encompass a particular weight
$1-\epsilon/\operatorname{poly}(n)$ is thus given by inverting~\eqref{eq:weight-of-region-Rf}. This may either
be done directly from the numerical histogram points, or from a fit model. This gives us a
region with high weight with respect to $\mu_{B^n}\left(\rho\right)$.

The method of Christandl and Renner may now be used to upgrade these regions to confidence
regions. Choose a confidence level $1-\epsilon$, and calculate the corresponding
$\operatorname{poly}(n)$ and $\delta$ as given in ref.~\cite{Christandl2012_Tomo}.  Recall
that a region with weight $1-\epsilon/\operatorname{poly}(n)$, once enlarged by $\delta$
in purified distance, is a confidence region with confidence $1-\epsilon$.

\begin{figure}
  \centering
  \includegraphics[width=\columnwidth]{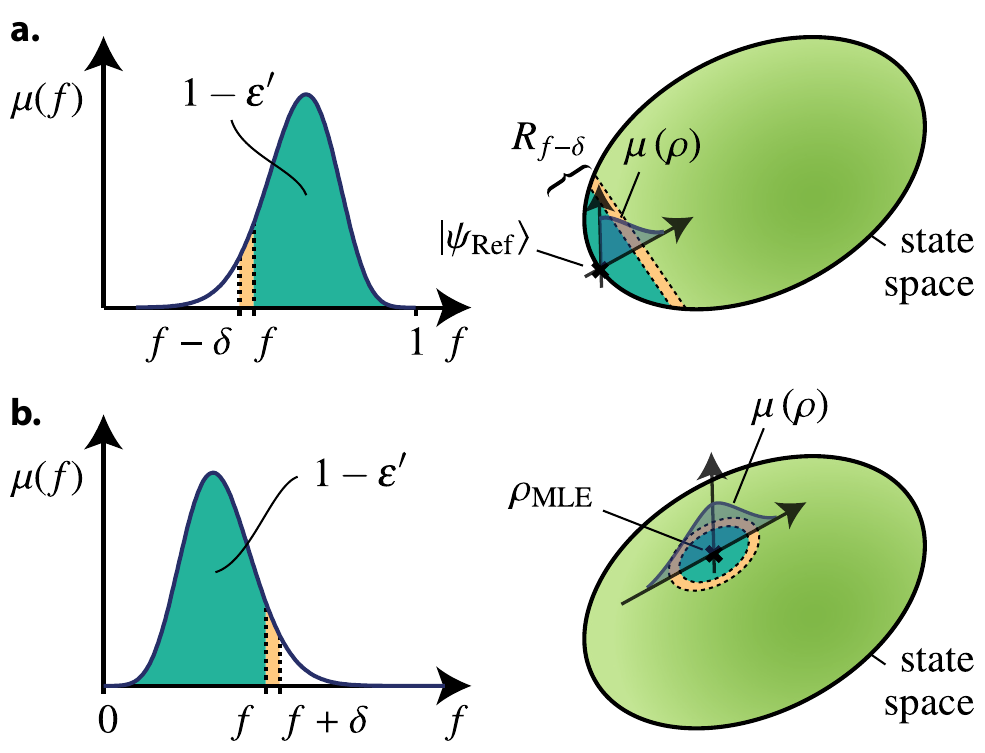}
  \caption{Construction of confidence regions from the histogram of the figure of
    merit. High weight intervals with respect to $\mu\left(f\right)$ in the histogram
    plots (left plots) correspond to high weight regions in state space with respect to
    $\mu_{B^n}\left(\rho\right)$ (right diagrams). An appropriate enlargement by a
    parameter $\delta$ yields confidence regions.  \fmtcapsubfig[a.] The case of the squared
    fidelity to a pure target state $\ket{\psi_\mathrm{Ref}}$ as figure of merit, i.e.\@
    $f\left(\rho\right) = F^2\left(\rho,\protect\proj{\psi_\mathrm{Ref}}\right) =
    \protect\matrixel{\psi_\mathrm{Ref}}{\rho}{\psi_\mathrm{Ref}}$.
    An interval $[f,1]$ in the histogram plot (turquoise region, left) corresponds to a
    region $R_f$ in state space consisting of all states with squared fidelity to
    $\ket{\psi_\mathrm{Ref}}$ higher than $f$ (turquoise region, right).  To construct a
    confidence region with a given confidence level $1-\epsilon$, first calculate
    $\epsilon/\operatorname{poly}(n)$ and $\delta$ as given in~\cite{Christandl2012_Tomo} and choose $f$ such
    that the turquoise shaded area in the histogram plot is $1-\epsilon/\operatorname{poly}(n)$. This means that
    the corresponding region $R_f$ has weight at least $1-\epsilon/\operatorname{poly}(n)$ under
    $\mu\left(\rho\right)$.  Then, the region $R_{f-\delta}$ consisting of all states with
    squared fidelity at least $f-\delta$ to $\ket{\psi_\mathrm{Ref}}$ form a confidence
    region of confidence level $1-\epsilon$ (turquoise and orange regions
    combined)~\cite{Christandl2012_Tomo}. The true state is almost surely in the region
    $R_{f-\delta}$; equivalently, the true fidelity to $\ket{\psi_\mathrm{Ref}}$ is almost
    surely better than $f-\delta$.  Observe that the regions constructed this way are
    linear slices of the quantum state space. This is because the squared fidelity to a
    pure state is linear. Also, the function $\mu\left(\rho\right)$ behaves approximately
    like a Gaussian around the maximum likelihood estimate $\rho_\mathrm{MLE}$; the
    illustration of $\mu\left(\rho\right)$ here corresponds to the case where
    $\rho_\mathrm{MLE}$ coincides with $\protect\proj{\psi_\mathrm{Ref}}$.
    \fmtcapsubfig[b.] The analogous construction applied to the case where the figure of merit
    is the trace distance to the maximum likelihood estimate $\rho_\mathrm{MLE}$. Here the
    regions $R_f$ and $R_{f+\delta}$ are trace distance balls around
    $\rho_\mathrm{MLE}$. If $R_f$ has weight at least $1-\epsilon/\operatorname{poly}(n)$, then $R_{f+\delta}$
    is a confidence region of confidence level $1-\epsilon$.}
  \label{fig:ConfRegionFigureOfMerit}
\end{figure}

In general, the $\delta$-enlargement can be translated into a cost in the corresponding bounding
figure of merit $f$. We'll derive here this cost for our particular cases of interest of
the fidelity to a pure reference state, the expectation value of an observable and the
trace distance to any reference state.  Consider first the case where the figure of merit
corresponds to the trace distance to a reference state $\rho_\mathrm{Ref}$,
$f\left(\rho\right)=D\left(\rho,\rho_\mathrm{Ref}\right) =
\frac12\norm{\rho-\rho_\mathrm{Ref}}_1$.  Note the reverse inequality is used
in~\eqref{eq:def-R-f-from-histogram}.  Then, consider the region $R_{f+\delta}$. Now,
we'll see that $R_{f+\delta}$ contains the set $R_f$ enlarged by $\delta$ in purified
distance. Indeed, if $\rho\in R_f$, and $\sigma$ is such that
$P\left(\rho,\sigma\right)=\sqrt{1-F^2\left(\rho,\sigma\right)}\leqslant\delta$, then we
may use the triangle inequality, along with the fact that
$D\left(\rho,\sigma\right)\leqslant
P\left(\rho,\sigma\right)$~\cite{Tomamichel2010IEEE_Duality}, to see that
$D\left(\sigma,\rho_\mathrm{Ref}\right)\leqslant f+\delta$, and deduce that $\sigma\in
R_{f+\delta}$. Thus, if $R_f$ is a region with weight $1-\epsilon/\operatorname{poly}(n)$, then $R_{f+\delta}$
is a confidence region of confidence level at least $1-\epsilon$. This construction is
depicted graphically in \autoref{fig:ConfRegionFigureOfMerit}.
Of course, the same reasoning applies to the case where $f\left(\rho\right) =
P\left(\rho,\rho_\mathrm{Ref}\right)$ is the purified distance to a reference state.

Consider also the case where $f\left(\rho\right)=\tr\left(\rho W\right)$ is the
expectation of an observable $W$. First, assume that the reverse inequality direction is
used in~\eqref{eq:def-R-f-from-histogram}. Then the $\delta$-enlargement of $R_f$ is
included in the region $R_{f+w\delta}$, where we've assumed that the eigenvalues $z_j$ of
$W$ lie within an interval of size $w$, i.e.\@ $w_-\leqslant z_j \leqslant w_+$ and
$w=w_+-w_-$, or equivalently, $w_-\Ident\leqslant W\leqslant w_+\Ident$ and
$w=w_+-w_-$. Indeed, assume $f\left(\rho\right)\leqslant f$ and
$P\left(\rho,\sigma\right)\leqslant\delta$. Then
$D\left(\rho,\sigma\right)\leqslant P\left(\rho,\sigma\right)$ and by properties of the
trace distance there exists $\Delta_\pm\geqslant0$ such that
$\sigma-\rho=\Delta_+-\Delta_-$ and
$\frac12\tr\left(\Delta_++\Delta_-\right)=\tr\Delta_+=\tr\Delta_- =
D\left(\rho,\sigma\right)\leqslant\delta$~\cite{BookWilde2013QIT}.
Now,
$f\left(\sigma\right) = \tr\left(\sigma W\right) = f\left(\rho\right) +
\tr\left(\Delta_+W\right) - \tr\left(\Delta_-W\right) \leqslant f + w_+\tr\Delta_+ -
w_-\tr\Delta_- \leqslant f+w\delta$,
and $\sigma\in R_{f+w\delta}$.  If the forward direction inequality is used
in~\eqref{eq:def-R-f-from-histogram} instead of the reverse, then the same argument above
is easily adapted to show that the $\delta$-enlargement of $R_f$ is included in the region
$R_{f-w\delta}$, where $w$ is defined in the same way.

Note also that the case of the squared fidelity to a pure reference state
$\ket{\psi_\mathrm{ref}}$ is given by
$f\left(\rho\right) = F^2\left(\rho,\proj{\psi_\mathrm{Ref}}\right) =
\matrixel{\psi_\mathrm{Ref}}{\rho}{\psi_\mathrm{Ref}}$,
and is thus the expectation value of the observable $\proj{\psi_\mathrm{Ref}}$. More
precisely, we have $w_+=1$, $w_-=0$ and $w=1$, and the inequality direction used
in~\eqref{eq:def-R-f-from-histogram} encompasses larger values for the fidelity in the
region; the enlarged set to consider is then simply $R_{f-\delta}$.

Remark that the error bars on the numerical estimate of $\mu\left(f\right)$, or on the
relevant fit parameters, should \emph{a priori} be propagated to the obtained confidence
regions. However since $\mu\left(f\right)$ decays exponentially, small errors on the fit
paramterers will only have a negligible effect on the $f$ required to contain a given
weight $1-\epsilon/\operatorname{poly}(n)$ as given by~\eqref{eq:weight-of-region-Rf}. This is just like
classical error bars---error bars hardly need their own error bars.


The final confidence regions obtained are still generally unmeaningfully large. For
example, if we try to construct confidence regions for the two superconducting qubits and
choose, say, $\epsilon=5\%$, then we see that
$\epsilon/\operatorname{poly}\left(n\right) \sim 10^{-37}$. Yes, that's
small.\footnote{This figure even uses an improved polynomial factor
  $\operatorname{poly}\left(n\right)=2n^{(d^2-1)/2}$~\cite{ChristandlTomoNotesPoly} over
  the original one in Ref.~\cite{Christandl2012_Tomo}.}  The corresponding $f$ required
(see construction in \autoref{fig:ConfRegionFigureOfMerit}) is $f\approx 0.85$, and we
can calculate $\delta\sim 0.1$. The final confidence region comprises then all states with
a fidelity to the target state in the range $[0.75,1]$. This analysis is in itself not
very useful, as it is fair to claim the experiment achieves considerably better precision
than that (compare with \autoref{fig:AnalysisSuperconductingQubits} of the main
text).  The solution we propose is to provide a characterization of the full function
$\mu\left(f\right)$ in terms of few parameters, from which we know that one can \emph{in
  principle} construct confidence regions for any desired confidence level. This is very
much akin to the error bars reported for a usual classical physical quantity: these may
typically represent one standard deviation of a value which is assumed to be Gaussian
distributed.  A confidence region of high confidence level could then be much larger than
the reported error bar.  For example, a confidence level of one part in a million requires
a region size of 5 standard deviations (or ``5 sigma'').

\section{Application of the method to simulated measurements}
\label{appx:application-simulated-experiment}
We have simulated measurements of individual Pauli operators on two qubits in the noisy
entangled state
\begin{align}
  \label{eq:rho-simulated-true-state}
  \rho = 0.95\,\proj\Psi + 0.05\,\frac{\Ident}{4}\ ,
\end{align}
with the pure entangled state $\ket\Psi = \left(\ket{01}+i\ket{10}\right)/\sqrt2$. Each
measurement setting consists of a pair of Pauli operators and has four outcomes, with a
total of 9 measurement settings.  Each setting was repeated $500$ times, resulting in
$4500$ total measurement outcomes. Our procedure yields histograms corresponding to three
different figures of merit (\autoref{fig:ExampleSimulatedData1}):
\begin{figure*}
  \centering
  \mysubfig\label{fig:ExampleSimulatedData1-f2tgt}%
  \mysubfig\label{fig:ExampleSimulatedData1-wit}%
  \mysubfig\label{fig:ExampleSimulatedData1-td}%
  \includegraphics{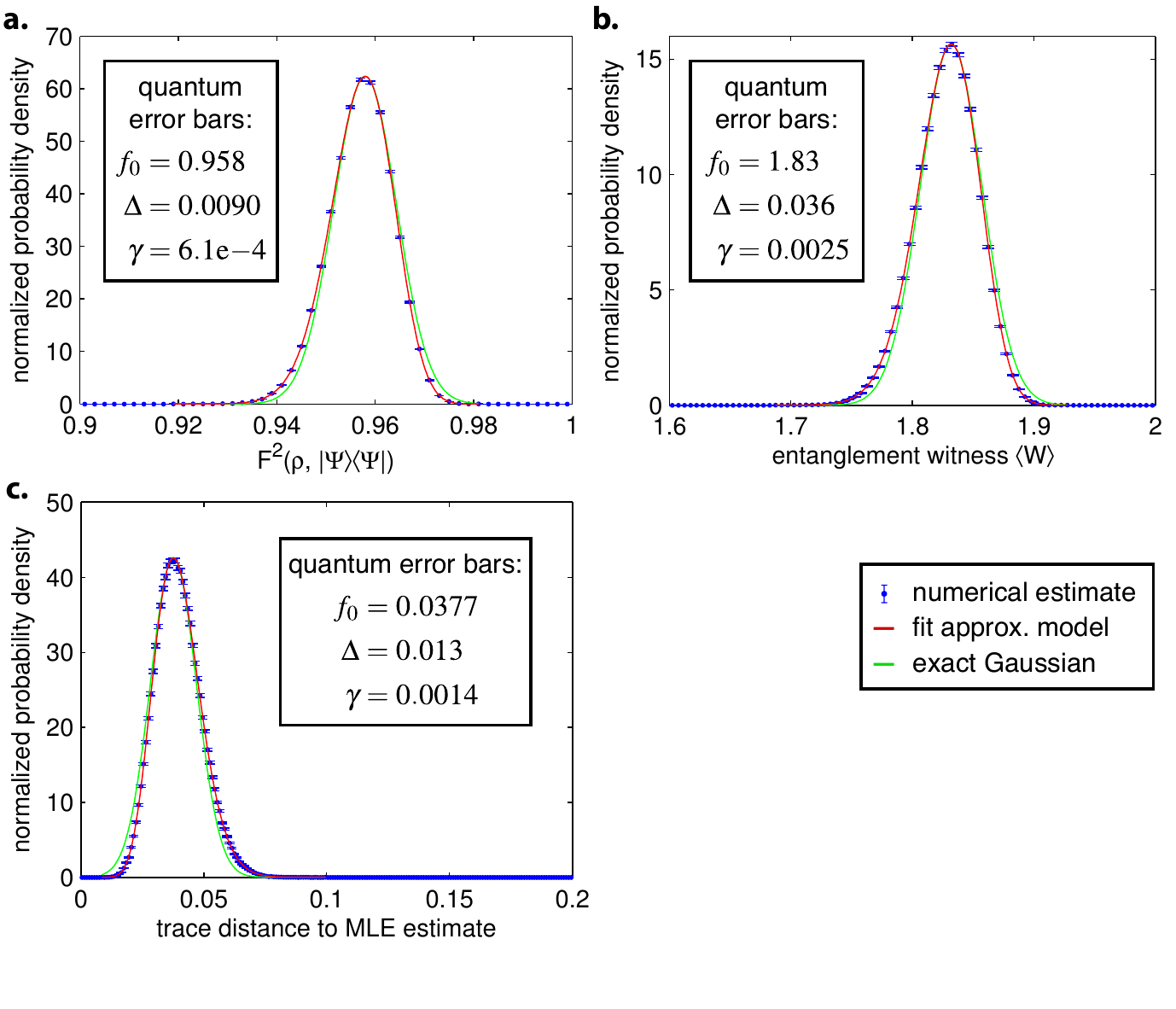}
  \caption{Simulated example data analyzed using our procedure, along with the quantum
    error bars. Two qubits in a noisy
    entangled state were measured with individual Pauli operators. \fmtcapsubfig[a.]
    Histogram of the squared fidelity to the target state $\ket\Psi$ under the
    distribution which is relevant for the construction of confidence regions using the
    method of Christandl and Renner~\cite{Christandl2012_Tomo}. The numerical estimate
    (blue points) were obtained using our procedure based on a Metropolis-Hastings random
    walk. The theoretical model fits the numerics well (red curve).  The quantum error
    bars characterize the distinctive features of the curve, which is interpreted as
    applying a ``skewing'' operation on an
    exact Gaussian (green curve).   Confidence regions may
    be constructed from this histogram by choosing regions with states that have a certain
    minimum fidelity to the target state, chosen such that the set has high weight with
    respect to this distribution. The true value of the squared fidelity of the state from
    which we have simulated measurements is in fact $0.963$; the shift is due to the
    increasing volume factor towards lower fidelity values. \fmtcapsubfig[b.] The same
    analysis is applied to the case of the expectation value of an entanglement
    witness. The witness is chosen to have positive expectation value only for entangled
    states, with a maximum at the maximally entangled state $\ket\Psi$ where
    $\protect\matrixel{\Psi}{W}{\Psi}=2$. \fmtcapsubfig[c.] The same analysis is again
    repeated for the case where the figure of merit is the trace distance to the maximum
    likelihood estimate.
}
  \label{fig:ExampleSimulatedData1}%
\end{figure*}
(a) the fidelity to the state $\ket\Psi$, (b) the expectation value of an entanglement
witness, and (c) the trace distance to the maximum likelihood estimate. The histograms
were each generated with one random walk instance for each of the 12 CPU cores
available. Each random walk produced $32768$ samples, yielding a total of
$12\times 32768=393216$ recorded samples. Error bars were obtained by binning analysis for
each run and combined with standard propagation of error bars. This analysis runs fast for
two qubits and can usually be obtained within minutes on usual hardware.

For the figure of merit (a), we have
$f\left(\rho\right) = F^2\left(\rho,\proj\Psi\right) = \matrixel{\Psi}{\rho}{\Psi}$. This
figure of merit is often used to report the accuracy of experimental preparations of
quantum states. Here the true value of this figure of merit is
$F^2\left(\rho,\proj\Psi\right)=0.9625$, given by the ``true
state''~\eqref{eq:rho-simulated-true-state} we used to simulate measurement outcomes.
In (b), the figure of merit is $f\left(\rho\right)=\tr\left(\rho W\right)$, with the
entanglement witness
$W=-\Ident-\sigma_X\otimes\sigma_Y+\sigma_Y\otimes\sigma_X-\sigma_Z\otimes\sigma_Z$. The
operator $W$ is chosen such that $\tr\left(\rho W\right)\leqslant0$ for all states $\rho$
which are not entangled, but also such that $\tr\left[\proj\Psi W\right]=2$.
For the last case considered, (c), we first calculate the maximum likelihood estimate
$\rho_\mathrm{MLE}$, and then define our figure of merit as
$f\left(\rho\right) = D\left(\rho,\rho_\mathrm{MLE}\right)$, where
$D\left(\rho,\sigma\right) := \frac12\norm{\rho-\sigma}_1$ is the trace distance. The
eigenvalues of the density matrix $\rho_\mathrm{MLE}$ are
$\left(0.000, 0.0105, 0.0240, 0.9655\right)$, meaning that the state lies on the border of
state space.

Observe that the peak maxima in \autoref{fig:ExampleSimulatedData1} do not correspond to
the values of the maximum likelihood estimate $\rho_\mathrm{MLE}$, even though the latter
is precisely the point where $\mu_{B^n}\left(\rho\right)$ is maximal by definition. This
is because of this increasing volume factor which shifts the peak.  Indeed, at
$\rho_\mathrm{MLE}$ we can evaluate
$F^2\left(\rho_\mathrm{MLE},\proj\psi\right) \approx 0.965$; the peak maximum in
\autoref{fig:ExampleSimulatedData1-f2tgt} is clearly shifted.

\begin{figure}
  \centering
  \includegraphics[width=\columnwidth]{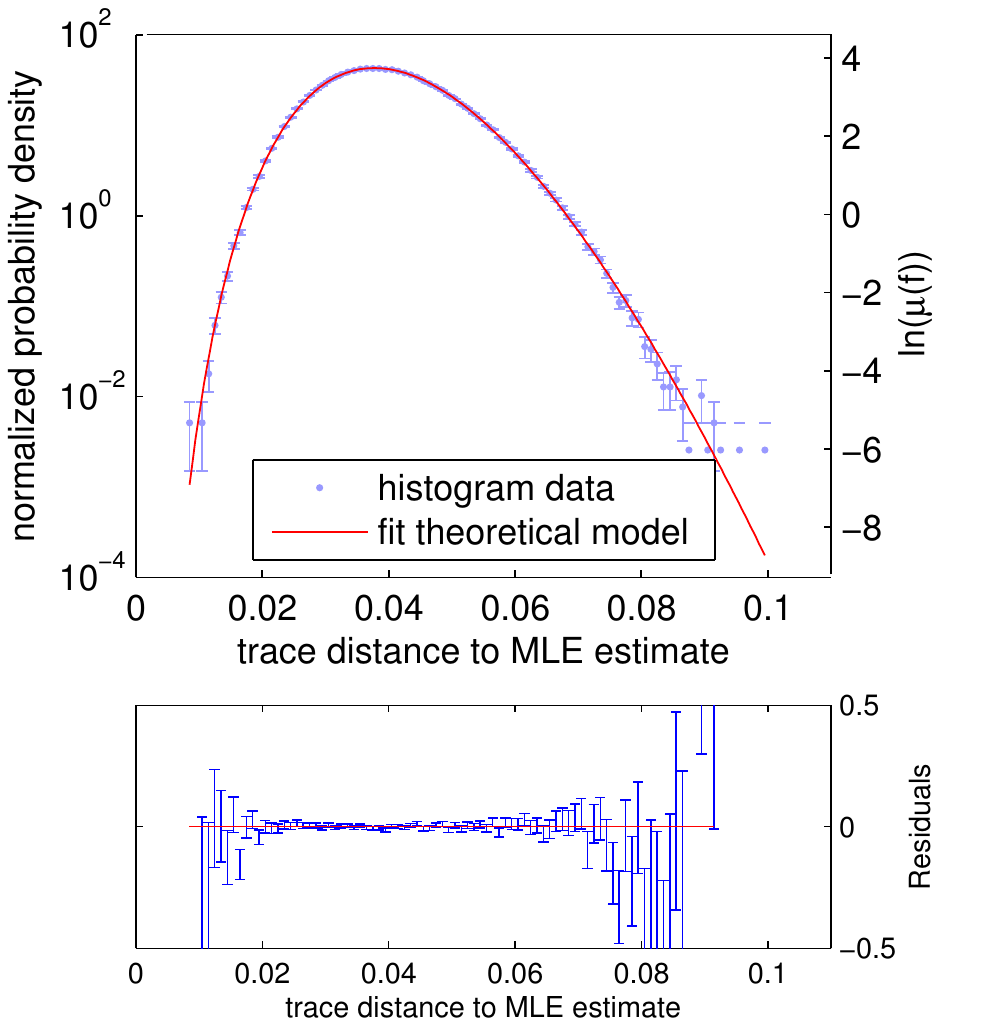}
  \caption{Fit of the logarithm of the numerically obtained histogram. The log scale
    allows to better appreciate the quality of the fit. The bottom plot shows the
    residuals of the fit for $\ln\mu\left(f\right)$, that is, the difference between the
    logarithm of the numerical histogram points and the fitted model for the function
    $\ln\mu\left(f\right)$. The low deviation from zero for the central points underscores
    the quality of our model (and that the error bars may be overestimated). The points at
    the sides correspond to regions with very low probability and where the numerical
    estimate is anyway expected to be unreliable.}
  \label{fig:exp06-figlogp-tomorun-config5td}
\end{figure}

Let us now apply the fit models to our numerical estimates. First, consider the trace
distance as figure of merit. The corresponding theoretical model
is~\eqref{eq:fit-model-logp-f-h} with $h=0$, as given in
\autoref{tab:TheoModelsFiguresOfMerit} of the main text:
$\ln\mu\left(f\right)\approx -a_2f^2-a_1f+m\ln f+c$. The model fits well to the numerical
estimate in \autoref{fig:ExampleSimulatedData1-td} (red curve). The fit was performed on
the logarithm of the histogram. We used weights for each point obtained by propagating the
error bars as
$\Delta\left[\ln\mu\right] = \abs{\partial\left(\ln\mu\right)/\partial\mu}\, \Delta\mu =
\Delta\mu/\mu$.
Points with obviously huge error bars were excluded from the fit. The raw fit for
$\ln\mu\left(f\right)$ is presented in \autoref{fig:exp06-figlogp-tomorun-config5td},
along with a plot of the residuals. The corresponding fit parameters are (with $95\%$
confidence bounds):
\begin{align}
  \begin{split}
    a_2 &= \makebox[3em][l]{722.8} (635.5, 810.1) \\
    a_1 &= \makebox[3em][l]{319.6} (305.6, 333.6) \\
    m   &= \makebox[3em][l]{14.09} (13.82, 14.36) \\
    c   &= \makebox[3em][l]{63.00} (61.71, 64.3).
  \end{split}
\end{align}
The value $m$ is close to $M-1=14$, which is the value we would predict
from~\eqref{eq:fit-model-p-calc-all-terms} without the $w\left[f,\Omega_0\right]$
term. This indicates that this latter volume factor is approximately constant in this
case.  The quantum error bars can be obtained with~\eqref{eq:QuantErrorBars},
\begin{align}
  \begin{split}
    f_0 &= 0.0377\ ; \\
    \Delta &= 0.013\ ;\\
    \gamma &= 0.0014\ .
  \end{split}
\end{align}

Our model also fits the numerically obtained histogram in
Figures~\ref{fig:ExampleSimulatedData1-f2tgt} and~\ref{fig:ExampleSimulatedData1-wit}
well. For \autoref{fig:ExampleSimulatedData1-f2tgt}, the appropriate fit model is
$\ln\mu\left(f\right) \approx -a_2\left(1-f\right)^2 - a_1\left(1-f\right) +
m\ln\left(1-f\right) + c$,
and for \autoref{fig:ExampleSimulatedData1-wit} we have used the model
$\ln\mu\left(f\right) \approx -a_2\left(2-f\right)^2 - a_1\left(2-f\right) +
m\ln\left(2-f\right) + c$,
as specified by \autoref{tab:TheoModelsFiguresOfMerit} of the main text.  The
respective quantum error bars are indicated on the corresponding plots in
\autoref{fig:ExampleSimulatedData1}.

\section{Application to experimental data and modeling noisy measurements}
\label{appx:application-experiment-superconducting}
As an illustration, we apply our method to analyze measurement data obtained for two
superconducting qubits in a Bell state prepared using the setup reported
in~\cite{Steffen2013Nat_deterministic}. The data were kindly provided by the authors of
Ref.~\cite{Steffen2013Nat_deterministic}. The measurement on an individual qubit is
carried out by a transmission measurement on a resonator coupled to that
qubit~\cite{Bianchetti2009PRA}. This measurement yields a random real value $I$ which is
distributed differently whether the qubit is in the $\ket0$ state or in the $\ket1$
state. Single-shot readouts are possible to reasonable accuracy using a simple threshold,
because the two distributions of $I$ corresponding to $\ket0$ and $\ket1$ have almost
non-overlapping support~\cite{Steffen2013Nat_deterministic}. However, we choose to model
the measurement process more precisely, as our method assumes the POVM effects correctly
incorporate any noise introduced by the measurement device itself. Here, we model the
measurement process as a real-valued POVM. A calibration measurement yields the
distributions $q_0(I)$ and $q_1(I)$ for trusted preparations of the $\ket0$ and $\ket1$
states respectively.  The measurement of the Pauli operator $\sigma_i$ is performed by
applying the appropriate unitary gate $U_i$ with high fidelity to bring the measurement
basis onto the computational basis. The effects corresponding to the real-valued POVM
including the rotation with $U_i$ are then
\begin{align}
  Q_i\left(I\right) = U_i^\dagger\begin{pmatrix} q_0\left(I\right) & 0\\
    0 & q_1\left(I\right) \end{pmatrix} U_i\ .
\end{align}
(We have ignored here errors in implementing the gate $U_i$.)  We could have used these
POVM effects directly for each measured value for each qubit in the expression for the
loglikelihood given by Eq.~(\ref*{eq:log-lambda-general}) of the main text,
however for practical purposes (to reduce the number of different POVM effects), we have
coarse-grained the values $I$ into $20$ different bins, yielding the discrete
distributions $q_0'\left(k\right)$ and $q_1'\left(k\right)$ for bin number $k$. In other
words, if the measured value $I$ is in bin number $k$, then the corresponding POVM effect
is
\begin{align}
  Q_i'\left(k\right) = U_i^\dagger\begin{pmatrix} q_0'\left(k\right) & 0\\
    0 & q_1'\left(k\right) \end{pmatrix} U_i\ .
\end{align}

The joint POVM effect corresponding to combining individual measurements on the two qubits
is simply given by tensoring the two POVM effects. For example, if the value $I_A$ is
measured on qubit $A$ (falling in bin $k_A$) and the value $I_B$ is measured on qubit $B$
(which falls in the bin $k_B$), then the joint POVM effect is simply
\begin{align}
  Q_{ij}'\left(k_A,k_B\right) = Q_i'\left(k_A\right) \otimes Q_j'\left(k_B\right)\ ,
\end{align}
where $U_i$ (resp.\@ $U_j$) is the rotation applied to qubit $A$ (resp.\@ qubit $B$)
before measuring the qubit in the computational basis.

We have analyzed the measurement data using the procedure described above. There were in
total $n=55\,677$ measurements. The histogram corresponding to the squared
fidelity to the target Bell state is depicted in
\autoref{fig:AnalysisSuperconductingQubits} of the main text.
Our theoretical model~\eqref{eq:fit-model-logp-h-f} (with $h=1$) fits the
numerical estimate well. The fit parameters are (with 95\% confidence bounds),
\begin{align}
  \begin{split}
    a_2 &= \makebox[4em][r]{$8511$}\quad (7909, 9112) \\
    a_1 &= \makebox[4em][r]{$-476.8$}\quad (-634.8, -318.7) \\
    m   &= \makebox[4em][r]{$42.53$}\quad (37.36, 47.69) \\
    c   &= \makebox[4em][r]{$125.4$}\quad (103.5, 147.2)\ .
  \end{split}
\end{align}
From these fit parameters, we finally derive the quantum error bars
\begin{align}
  \begin{split}
    f_0 &= 0.934 \ ; \\
    \Delta &= 0.0086\ ; \\
    \gamma &= 1.4\times10^{-4}\ .
  \end{split}
\end{align}

\section{Comparison with error bars from other methods}
\label{appx:compare-with-bootstrap}
\begin{figure}
  \centering
  \includegraphics[width=\columnwidth]{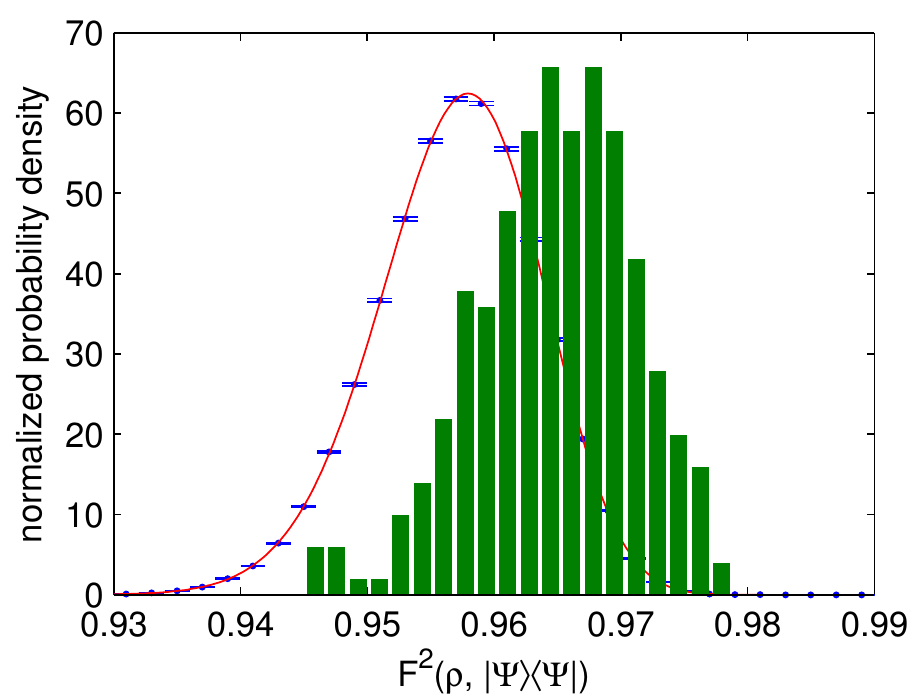}
  \caption{Comparison with error bars from existing bootstrapping methods. The blue points
    and red curve reproduce the numerical estimate and fit in
    \autoref{fig:ExampleSimulatedData1-f2tgt}. We have resampled random measurement
    outcomes from the maximum likelihood estimate and plotted the resulting reconstructed
    squared fidelities to the target state $\ket\Psi$ (green histogram bars). The bias of
    the curve from our method is due to the increasing volume factor in the direction of
    lower fidelity values. We see here that our method yields error bars of the same order
    of magnitude as bootstrapping. However our error bars are well-justified, because they
    can serve to construct confidence regions for any confidence level.}
  \label{fig:CompareSimpleBootstrapping}
\end{figure}

A currently used \emph{ad hoc} technique for obtaining error bars is
bootstrapping~\cite{BookEfron1994Bootstrap,Mitchell2003PRL_diagnosis,Home2009Sci_iontraps,BlumeKohout2012arXiv_Tomo,Schwemmer2015PRL_systematic}.
In our simulated experiment above, we have performed a simple parametric bootstrapping
analysis for comparison. We have simulated new measurement outcomes from
$\rho_\mathrm{MLE}$, using the same amount of measurements and the same settings as for
the original simulated experiment. We repeated the procedure many times to obtain in total
300 new datasets. For each dataset, we have reconstructed the corresponding maximum
likelihood estimate, and determined its squared fidelity to the target state $\ket\Psi$.
The histogram of these values is presented in \autoref{fig:CompareSimpleBootstrapping},
compared with the result of our method in \autoref{fig:ExampleSimulatedData1-f2tgt}.  We
see that the width of the distribution is approximately the same. The bias of $\sim1\%$
squared fidelity between the two methods is due to the increasing volume factor picked up
by our method in the direction of decreasing fidelities. Our error bars, however, have the
robust operational meaning as a means to construct confidence regions.

\section{Discussion of the reliability of the method}
\label{appx:reliability}

In our work, we provide several levels of reliability statements. First, obviously if
$\mu\left(f\right)$ can be exactly determined, then our error analysis is perfectly
reliable, assuming the given measurement operators are accurate.  In practice though, it
is only possible to approximate $\mu\left(f\right)$ with numerical techniques. However
these methods are standard and well-tested, and come with reliable error
estimates~\cite{Ambegaokar2010AJP_estimating}.  Thus with minimal reasonable assumptions
an error analysis based on this approximation of $\mu\left(f\right)$ is also reliable.

Furthermore, we provide an approximate theoretical model with only three fit parameters
and which explains well the numerical estimate obtained.  The quality of the fit over many
examples studied by the authors not only
presents additional strong indication that the numerical estimate is faithful, but also
shows that the result admits a simple representation with few parameters. Recall that the
numerical method does not rely on the assumptions and approximations used to derive the
theoretical fit model.  Also, because of its form the fit model is relatively robust to
small uncertainties in the fit parameters.

Our approximate fit model might fail though to describe the distribution accurately in
some extreme cases, for example if too few measurements are taken (however examples with
$n\sim100$ total measurements were well fit).  The model is also known not to apply to
figures of merit such as the fidelity to a non-pure state, or more generally, figures of
merit which do not satisfy~\eqref{eq:assumption-property-f-affine-in-r}.  However, the fit
in these cases is usually of bad quality, especially in the region of the peak.  In
practice, it is sufficient to rely on goodness-of-fit measures or visual inspection of the
quality of the fit to assert its validity.

\section{Overview of our software}
\label{appx:software}
We are releasing a software suite which accomplishes our procedure in a wide range of
settings~\cite{TomographerCxx}. The project is composed of a program ready for use, which
is built upon a modular, generic C++ framework designed for flexibility and speed.

We expect our program to be directly usable in most experimental applications. Our program
takes as input a list of POVM effects $E_k$, which are assumed to be independent, and a
list of frequencies $n_k$ which indicate how many times each corresponding POVM effect was
observed. Further inputs include settings for the histogram range and number of bins,
which figure of merit to use, parameters of the Metropolis-Hastings random walk, the
number of times to repeat the random walk, the error analysis method, etc.  The output of
the program is the histogram as displayed for example in
\autoref{fig:ExampleSimulatedData1}, as well as
\autoref{fig:AnalysisSuperconductingQubits} of the main text, with corresponding
error bars.  We refer to the project's hosted location~\cite{TomographerCxx} for further
documentation and detailed information about its usage. The histograms presented in this
work were all obtained using our software.

This program is itself built upon a generic C++ framework with a collection of tools which
may be used to specialize our method to more complex setups. We provide for example tools
to specify the data for a quantum tomography problem, an implementation of an abstract
Metropolis-Hastings random walk, an interface to collect statistics during this random
walk, as well as tools for parallel processing several random walk instances. The code is
written using a technique called C++ template
metaprogramming~\cite{BookAlexandrescu2001ModernCxxDesign}, which allows to write generic
code which is flexible and reusable, but which at compile-time is translated into highly
optimized low-level machine instructions. Our project relies on the Eigen and Boost
libraries~\cite{EigenCxx,BoostCxx}, in particular for linear algebra calculations.  These
libraries also make extensive use of this technique.

Some tasks are not covered by our program. If required by the figure of merit, finding the
maximum likelihood estimate $\rho_\mathrm{MLE}$ can be accomplished by minimizing the
loglikelihood $\lambda\left(\rho\right)$. Since this function is convex, the solution can
be found efficiently. In most of our examples, we used CVX, a MATLAB package for
specifying and solving convex problems~\cite{Grant2008_convex,Grant2014_soft_cvx}.  Also,
in order to determine the fit parameters corresponding to the histogram, we resorted
to MATLAB's curve fitting toolbox.  A MATLAB script is provided along with the software to
ease this task, and to calculate the quantum error bars.


Our program is currently limited to POVM effects with a product structure. However, this
is generically the case, even e.g.\@ for adaptive
tomography~\cite{Teo2011PRA,Sugiyama2012PRA_AdaptiveTomo,Mahler2013PRL_adaptive}.  We have
successfully used our code to analyze simulated measurements of Pauli operators of up to
at least 5 qubits on our hardware, and we expect further improvements will increase this
limit.



\bibliography{\jobname.bibolamazi}

\end{document}